\documentclass[12pt]{article}
\usepackage[dvips]{color}
\usepackage{epsfig}
\usepackage{amsmath}
\usepackage{graphicx}

\begin{document}
\title{ Thermodynamics of a Sufficient Small Singly Spinning Kerr-AdS Black Hole}
\author{{Behnam Pourhassan$^{a,}$\thanks{Email: b.pourhassan@du.ac.ir}, Mir Faizal$^{b,c,}$\thanks{Email: mirfaizalmir@gmail.com}}\\
$^{a,}${\small {\em School of Physics, Damghan University, Damghan, Iran}}\\
$^{b,}${\small {\em  Irving K. Barber School of Arts and Sciences,  University of British Columbia - Okanagan,}}\\
{\small {\em Kelowna,   BC V1V 1V7, Canada}}
\\
$^{c,}${\small {\em  Department of Physics and Astronomy, University of Lethbridge,}}\\
{\small {\em Lethbridge, AB T1K 3M4, Canada}}
}
\date{}
\maketitle

\begin{abstract}
In this paper, we will analyze the thermodynamics of a small singly spinning Kerr-AdS black hole. As the black hole will be sufficient small,
its temperature will be large and so we can not neglect the effects of thermal fluctuations.
We will demonstrate that  these thermal fluctuations correct the
entropy of singly spinning Kerr-AdS black hole by a logarithmic correction term. We will analyze the implications of the logarithmic correction
on other thermodynamic properties of this black hole, and analyze the stability of such a black hole. We will observe that this form of correction
becomes important when
the size of the black hole is sufficient small. We will also analyze the effect of these thermal
fluctuations on the critical phenomena for such a black hole.
{\bf Pacs:} 04.50.Gh\\\\
\end{abstract}

\section{Introduction}
Black holes are maximum entropy objects, and so a black hole has more entropy than any other object with the same volume \cite{1, 1a, 2, 4, 4a}.
It is important to associate a maximum entropy with a black hole as finite entropy objects
can cross the horizon of a black hole. This would spontaneously reduce the entropy of the universe, and thus
the second law of thermodynamics can get violated if a maximum entropy is not associated with a black hole.
This maximum entropy associated with a black hole  scales with the area of the horizon. In fact, the entropy of the black hole can be
expressed in terms of the area of the horizon as $s = A/4$.  The observation that the maximum entropy of a region of space scales with its area
has motivated the holographic principle \cite{5, 5a}. This principle states that the  number of   degrees of freedom
in any region of space is  equal to the number of degrees of freedom on the boundary of that region.\\
The holographic principle has found various applications in many different areas of physics. However, it is expected that
the holographic principle will get   modified near Planck scale \cite{6, 6a}.  This is because the area-entropy law of black holes
gets modified due to the quantum gravitational effects.   In fact, almost all approaches to quantum gravity  predict the same functional
form for these   quantum corrections to
the area-entropy relation i.e., the area-entropy law gets corrected by a  logarithmic correction term. However,
the coefficient of this logarithmic correction term depends on the specific approach chosen, and is different for different
approaches to quantum gravity. Such logarithmic correction term has been obtained using
 the  non-perturbative quantum  general
relativity  \cite{1z}. This was done by using the relation between the
density of states of a black hole and the conformal blocks of a well
defined  conformal field theory. The Cardy formula has  been used for obtaining such corrections terms
to the area-entropy relation \cite{card}.  The correction  for a BTZ black hole has been calculated, and it has been demonstrated  that
these  are  logarithmic  corrections \cite{card}.\\
The effect of matter fields surrounding a black hole has  been studied \cite{other, other0, other1}. This analysis has also been used to obtain corrections
to the area-entropy relation, and it was observed that this correction term is logarithmic.
The string theoretical corrections to the entropy of a black hole have been calculated,  and it has been found that the entropy of
a black hole gets corrected by  logarithmic term generated from string theoretical effects  \cite{solo1, solo2, solo4, solo5}.
The logarithmic correction to the entropy of a  dilatonic black hole has been obtained  \cite{jy}. The partition
function of a black hole  has  been used to obtain the  logarithmic correction to the area-entropy law of a black hole  \cite{bss}.
The corrections obtained using the generalized uncertainty principle  are also logarithmic \cite{mi, r1}. The thermodynamics and statistics of
G\"{o}del black hole with logarithmic correction has been   studied \cite{Pourdarvish}. Furthermore, $P-V$ criticality of dyonic charged AdS black
hole with a logarithmic correction has been also been analyzed \cite{Sadeghi}.\\
It may be noted that in the Jacobson formalism, the  Einstein's equation can be derived from the first law of thermodynamics \cite{z12j, jz12}.
In this formalism, it is required that the
Clausius relation holds for all the local Rindler causal horizons through each space-time point, and this gives rise to the Einstein's equation.
As there exists a relation between the geometry and  thermodynamics of a black hole, we expect that  thermal fluctuations in the thermodynamics will give
rise to the quantum fluctuations in the metric. In fact, it has been demonstrate that the entropy of the black hole gets corrected by
logarithmic terms due to these thermal fluctuations \cite{l1, SPR}. As such corrections are expected to occur from most approaches to quantum gravity,
we will study the effect of such term on our system. Furthermore, as the coefficient of such terms depends on the approach to
the quantum gravity, we will not fix the coefficient of such correction term, and hence see the effect of such correction term
on the system, which would be generated from different approaches to quantum gravity.\\
Already, the effect of thermal fluctuations on the thermodynamics quantities of an AdS charged black hole has been analyzed \cite{1503.07418}.
It was demonstrated that  the  thermal fluctuations  decrease the certain   thermodynamics potentials associated with the system,
for example, the  free energy reduced due to these fluctuations.
The modification to the thermodynamics of a black Saturn because of    the thermal fluctuations has also been studied   \cite{1505.02373},  and it was observed  that
logarithmic corrections do not affect stability of the black Saturn. The
logarithmic corrections to entropy of  a modified Hayward black hole has   been analyzed \cite{1603.01457},  and it was observed that the
value of the pressure and internal energy reduced due to
such corrections. It is also demonstrated that the first law of thermodynamics is satisfied for the
these black hole in the presence of thermal fluctuations. The correction to the thermodynamics of a
 charged dilatonic black Saturn have been studied   \cite{1605.00924}, and it has been demonstrated
that for this system the corrections obtained from a conformal field theory are the same as the corrections obtained from the fluctuations in the energy.
It may be noted that such corrections where studied by
analyzing the thermal fluctuations   very close to the equilibrium, and this approximation is expected to  breakdown near the Planck scale.
This is because near the Planck scale, the thermal fluctuations will become so large that the equilibrium thermodynamics cannot be used to describe the system.
However, as long as the system  remain close to the equilibrium,  it is possible to analyze  the effect of thermal fluctuations as a perturbation around
the equilibrium state   \cite{Landau, fl}.
\\
In this paper, we will analyze the effects of such thermal fluctuation on a higher dimensional singly spinning Kerr-AdS black hole \cite{1411.4309}. It may be
noted that the    thermodynamics of such a black hole has already been studied \cite{4a, 1510.00085},
and we shall analyze the corrections to the thermodynamics by thermal fluctuations. We will show that thermal fluctuations have an important effect
on the thermodynamics and
critical points of singly spinning Kerr-AdS black holes in higher dimensions. We will study the  effect of logarithmic correction on the partition function,
and also investigate the   special case of an ordinary Kerr-AdS black hole in four dimensions.

\section{Singly Spinning Kerr-AdS Black Hole}
In this section, we will discuss thermodynamic properties of a higher dimensional singly spinning Kerr-AdS black hole.
The metric for a  $d$-dimensional Kerr-AdS black hole  in Boyer-Lindquist coordinates can be written as \cite{Gibb},
\begin{eqnarray}\label{A1}
ds^{2}&=&-W\left(1+\frac{r^{2}}{l^{2}}\right)d\tau^{2}+\frac{2m}{U}\left(Wd\tau-
\sum_{i=1}^{N}\frac{a_{i}\mu_{i}^{2}d\varphi_{i}}{\Xi_{i}}\right)^{2}\nonumber\\
&+&\sum_{i=1}^{N}\frac{r^{2}+a_{i}^{2}}{\Xi_{i}}\mu_{i}^{2}d\varphi_{i}^{2}+\frac{U dr^{2}}{\mathcal{F}-2m}+\sum_{i=1}^{N+\varepsilon}\frac{r^{2}+a_{i}^{2}}{\Xi_{i}}d\mu_{i}^{2}\nonumber\\
&-&\frac{1}{W(r^{2}+l^{2})}\left(\sum_{i=1}^{N+\varepsilon}\frac{r^{2}+a_{i}^{2}}{\Xi_{i}}\mu_{i}d\mu_{i}\right)^{2},
\end{eqnarray}
where
\begin{eqnarray}\label{A2}
W&=&\sum_{i=1}^{N+\varepsilon}\frac{\mu_{i}^{2}}{\Xi_{i}},\nonumber\\
U&=&r^{\varepsilon}\sum_{i=1}^{N+\varepsilon}\frac{\mu_{i}^{2}}{r^{2}+a_{i}^{2}}\prod_{j}^{N}(r^{2}+a_{j}^{2}),\nonumber\\
\mathcal{F}&=&r^{\varepsilon-2}\left(1+\frac{r^{2}}{l^{2}}\right)\prod_{j}^{N}(r^{2}+a_{j}^{2}),\nonumber\\
\Xi_{i}&=&1-\frac{a_{i}^{2}}{l^{2}}.
\end{eqnarray}
Here, $m$ and $a_{i}$ are mass and rotation parameters, respectively. The coordinates $\mu_{i}$ satisfy the following constraint
\begin{equation}\label{A3}
\sum_{i=1}^{N+\varepsilon}\mu_{i}^{2}=1,
\end{equation}
where $\varepsilon=0$ for odd $d$ or $\varepsilon=1$ for even $d$. In the case of $d = 4$, the above space-time reduces to the four-dimensional Kerr-AdS metric.
Thermodynamics of this system has already been studied  \cite{4a}.  The mass of this
black hole  (as well as enthalpy) is given by,
\begin{equation}\label{A4}
M=\frac{m\omega_{d-2}}{4\pi(\prod_{j}\Xi_{j})}\left(\sum_{i=1}^{N}\frac{1}{\Xi_{i}}-\frac{1-\varepsilon}{2}\right),
\end{equation}
where $\omega_{d-2}$ is the volume of the unit $(d-2)$-sphere which is given by,
\begin{equation}\label{A5}
\omega_{d-2}=\frac{2\pi^{(\frac{d-1}{2})}}{\Gamma(\frac{d-1}{2})}.
\end{equation}
The angular momenta is  given by,
\begin{equation}\label{A6}
J_{i}=\frac{a_{i}m\omega_{d-2}}{4\pi\Xi_{i}(\prod_{j}\Xi_{j}))}.
\end{equation}
The angular velocities of the horizon is given by,
\begin{equation}\label{A7}
\Omega_{i}=\frac{a_{i}(r_{+}^{2}+l^{2})}{l^{2}(r_{+}^{2}+a_{i}^{2})}.
\end{equation}
Furthermore, the temperature $T$ can be expressed as
\begin{equation}\label{A8}
T=\frac{1}{2\pi}\left[r_{+}(1+\frac{r_{+}^{2}}{l^{2}})
\sum_{i=1}^{N}\frac{1}{r_{+}^{2}+a_{i}^{2}}-\frac{1}{r_{+}}\left(\frac{1}{2}-\frac{r_{+}^{2}}{2l^{2}}\right)^{\varepsilon}\right].
\end{equation}
The entropy of this black hole  is given by  $s=\frac{A}{4}$,  where
\begin{equation}\label{A9}
A=\frac{\omega_{d-2}}{r_{+}^{1-\varepsilon}}\prod_{i=1}^{N}\frac{r_{+}^{2}+a_{i}^{2}}{\Xi_{i}}.
\end{equation}
Here, the horizon radius $r_{+}$ is the largest root of $\mathcal{F} - 2m = 0$. The thermodynamic volume is given by,
\begin{equation}\label{A10}
V=\frac{Ar_{+}}{d-1}+\frac{8\pi}{(d-1)(d-2)}\sum_{i}a_{i}J_{i}.
\end{equation}
\\
In this paper, we  will    study the logarithmic corrections to the thermodynamics of singly spinning Kerr-AdS black hole. A
singly spinning Kerr-AdS black holes can be described using  one non-zero rotation parameter $a_{1}=a$ (while other $a_{i}$ are zero), and
so the metric (\ref{A1}) takes the following form,
\begin{eqnarray}\label{B1}
ds^{2}&=&-\frac{\Delta}{\rho^{2}}(dt-\frac{a}{\Xi}\sin^{2}\theta d\varphi)^{2}+\frac{\rho^{2}}{\Delta}dr^{2}+\frac{\rho^{2}}{\Sigma}d\theta^{2}\nonumber\\
&+&\frac{\Sigma\sin^{2}\theta}{\rho^{2}}[a dt-\frac{r^{2}+a^{2}}{\Xi}d\varphi]^{2}+r^{2}\cos^{2}\theta d\Omega_{d-4}^{2},
\end{eqnarray}
where
\begin{eqnarray}\label{B2}
\Delta&=&(r^{2}+a^{2})(1+\frac{r^{2}}{l^{2}})-2mr^{5-d},\nonumber\\
\Sigma&=&1-\frac{a^{2}}{l^{2}}\cos^{2}\theta,\nonumber\\
\rho^{2}&=&r^{2}+a^{2}\cos^{2}\theta,\nonumber\\
\Xi_{i}&=&1-\frac{a^{2}}{l^{2}}.
\end{eqnarray}
Thermodynamics quantities associated with this metric  have been studied   \cite{Gibb2}.  The
temperature and entropy of this black hole, are given by
\begin{equation}\label{B3}
T=\frac{1}{2\pi}\left[r_{+}(1+\frac{r_{+}^{2}}{l^{2}})\left(\frac{1}{r_{+}^{2}+a^{2}}+\frac{d-3}{2r_{+}^{2}}\right)-\frac{1}{r_{+}}\right],
\end{equation}
and
\begin{equation}\label{B4}
s=\frac{\omega_{d-2}}{4}\frac{(r_{+}^{2}+a^{2})r_{+}^{d-4}}{\Xi_{i}},
\end{equation}
where $r_{+}$ is the largest positive real root of $\Delta=0$, which has been obtained using from the first equation in (\ref{B2}).

\begin{figure}[h!]
 \begin{center}$
 \begin{array}{cccc}
\includegraphics[width=50 mm]{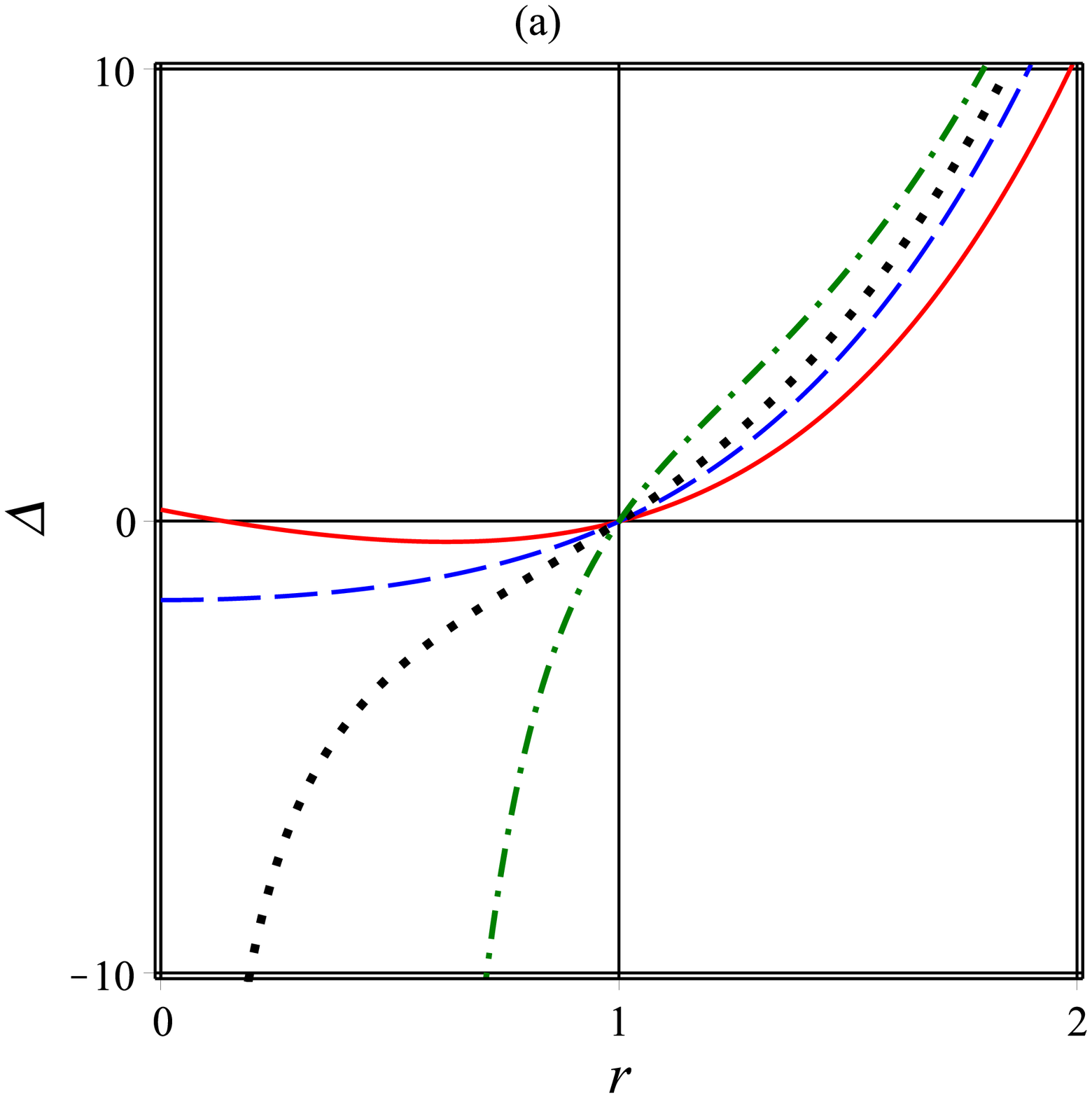}&\includegraphics[width=50 mm]{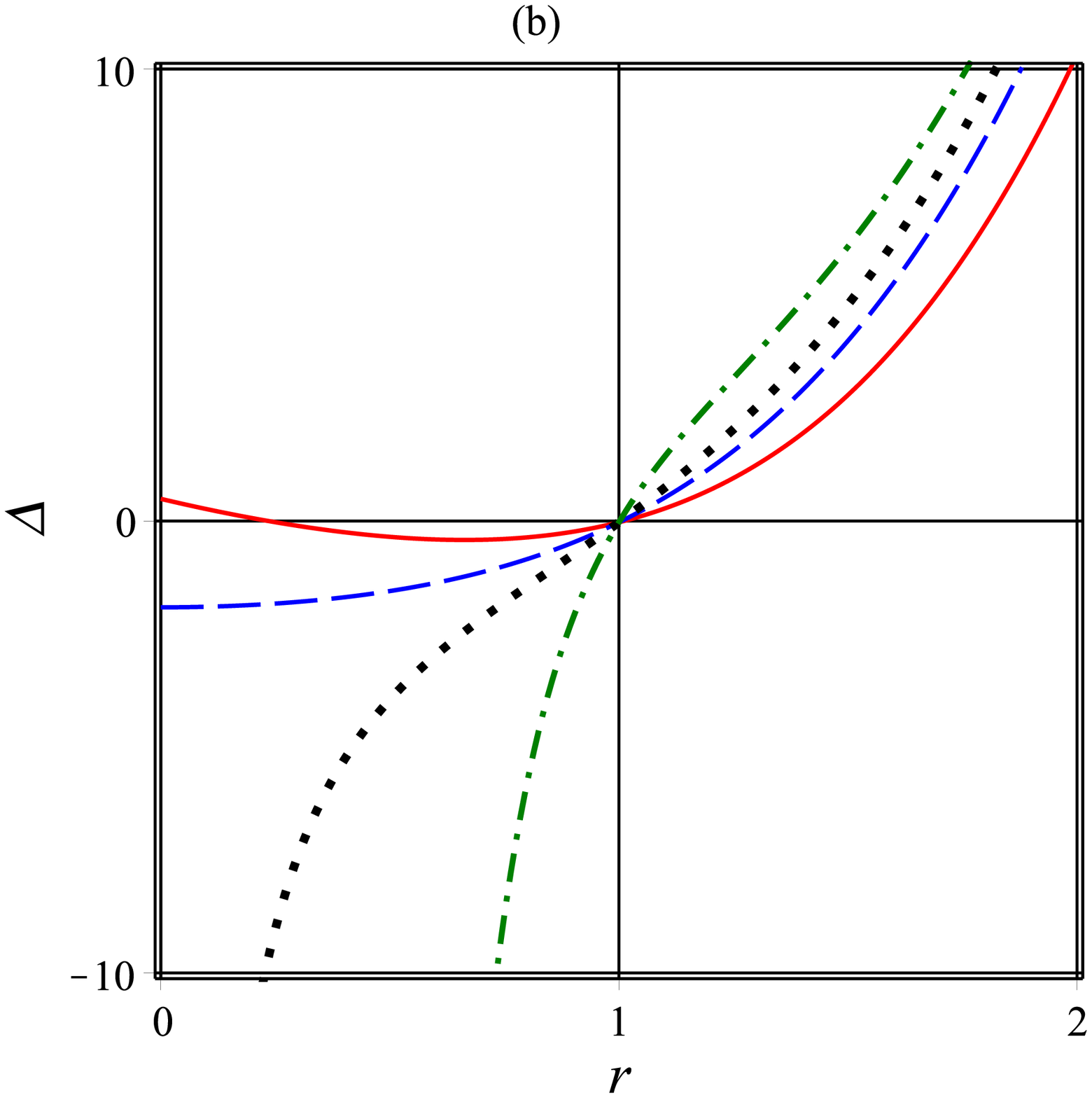}&\includegraphics[width=50 mm]{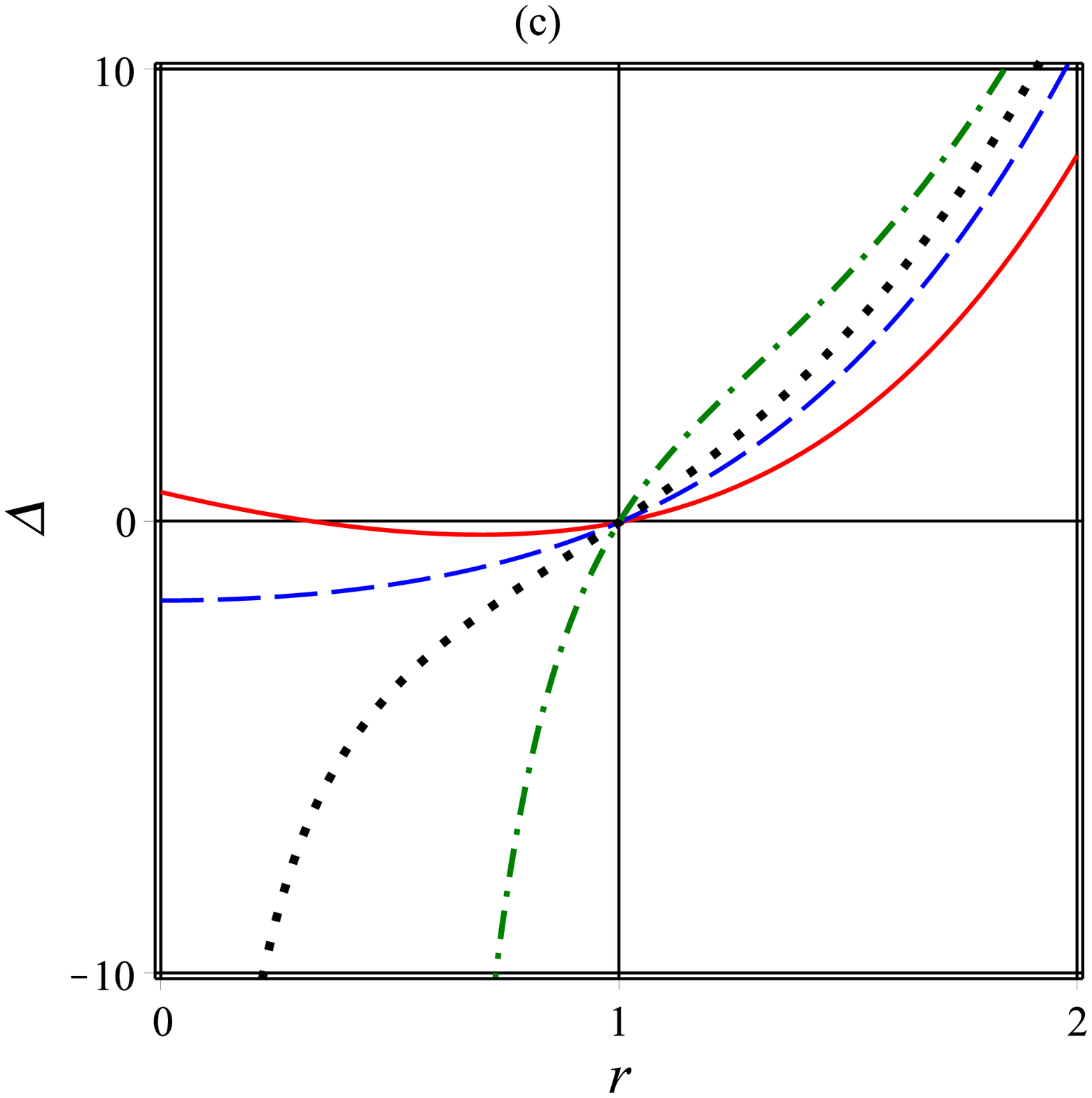}\\
\includegraphics[width=50 mm]{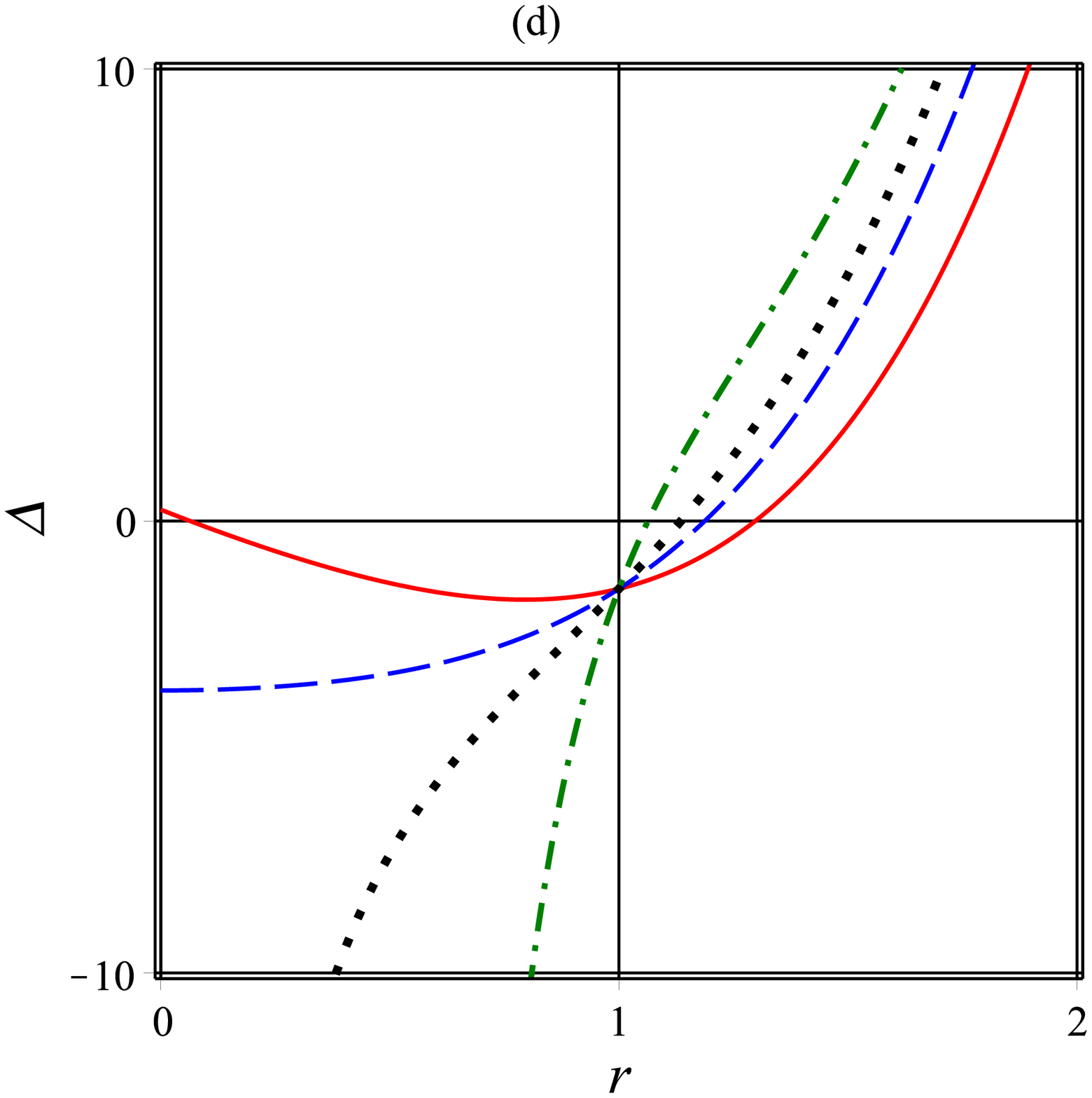}&\includegraphics[width=50 mm]{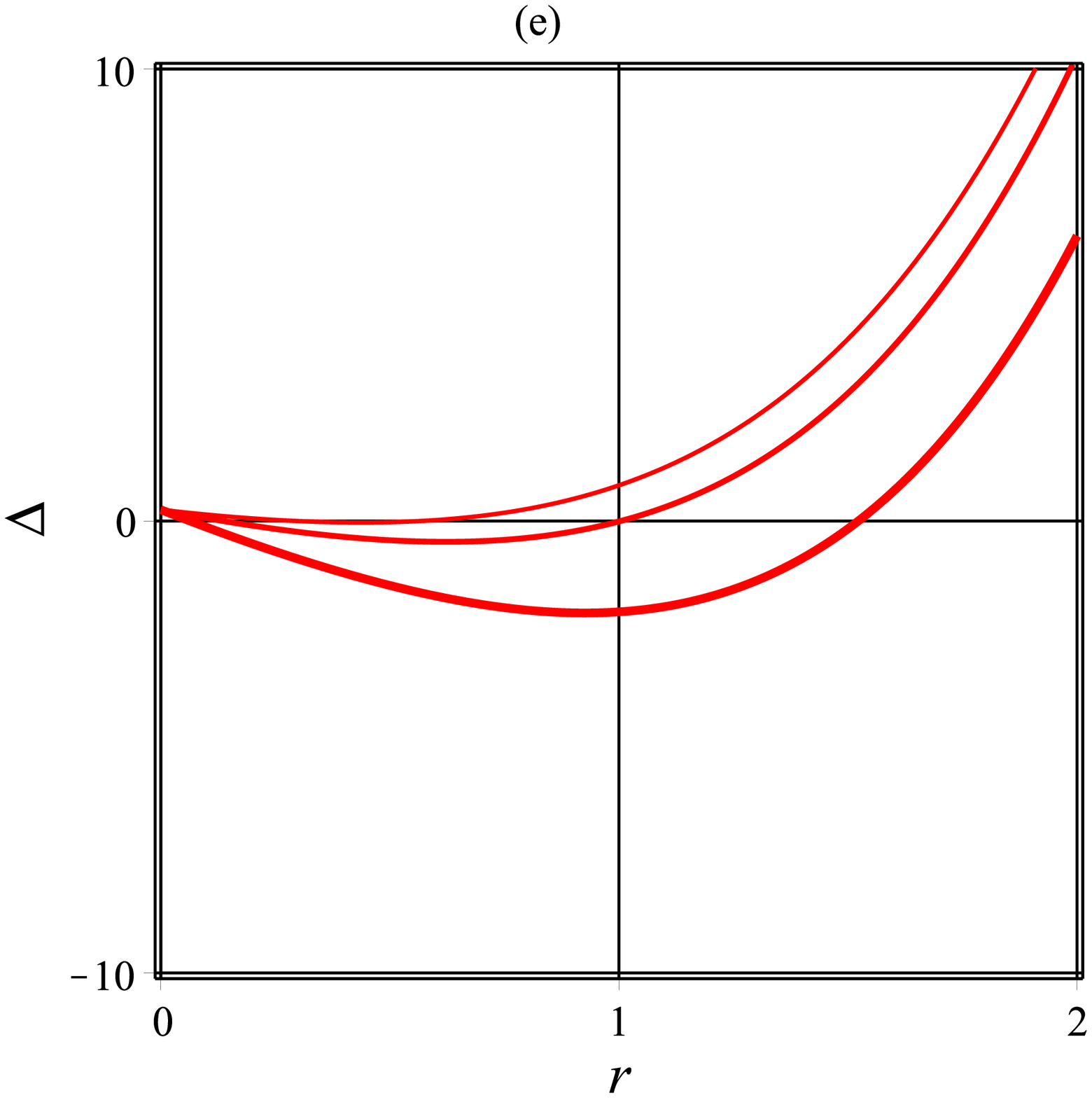}&\includegraphics[width=50 mm]{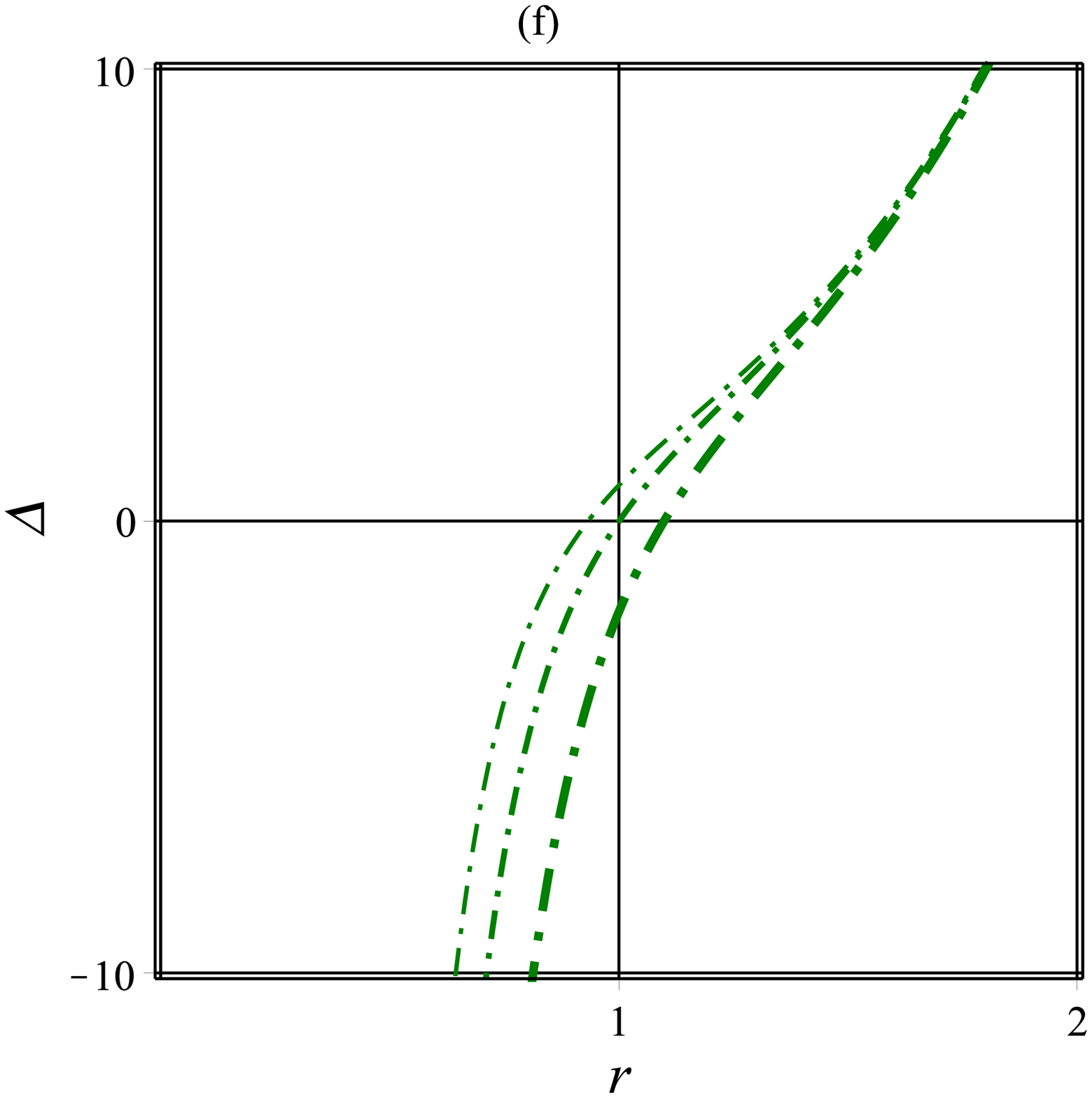}
 \end{array}$
 \end{center}
\caption{$\Delta$ in terms of $r$. (a) $l=1.3$, $m=1$, and $a=0.5$. (b) $l=1.3$, $m=1.2$, and $a=0.7$. (c) $l=1.5$, $m=1.2$, and $a=0.8$.
(d) $l=1$, $m=2$, and $a=0.5$. (e) $l=1.3$, $a=0.5$, $m=0.6$ (upper thin), $m=1$ (middle), and $m=2$ (lower thick). (f) $l=1.3$, $a=0.5$, $m=0.6$
(upper thin), $m=1$ (middle), and $m=2$ (lower thick). $d=4$ (solid red), $d=5$ (dashed blue), $d=6$ (dotted black), $d=10$ (dash dotted green).}
 \label{fig:1}
\end{figure}
\begin{figure}[h!]
 \begin{center}$
 \begin{array}{cccc}
\includegraphics[width=55 mm]{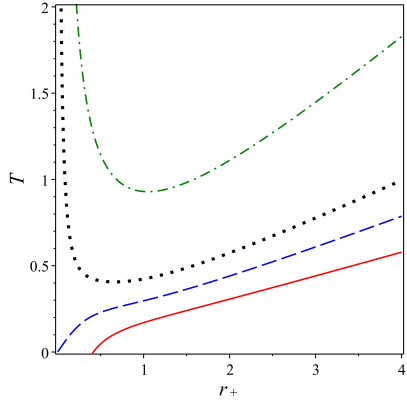}
 \end{array}$
 \end{center}
\caption{Temperature in terms of $r_{+}$ for $l=1.3$, and $a=0.5$. $d=4$ (solid red), $d=5$ (dashed blue), $d=6$ (dotted black), $d=10$ (dash dotted green).}
 \label{fig:2}
\end{figure}

Plots of the Fig. \ref{fig:1} shows that there is at least one positive root. There are special choices of $(l,m,a)$
parameters such as $(1.3,1,0.5)$, $(1.5,1.2,0.8)$, $(1.3, 1.2, 0.7)$,... where
$r_{+}=1$ for all $d$. We see small differences between Figs. \ref{fig:1} (a), (b) and (c)
where $r_{+}=1$, while Fig. \ref{fig:1} (d) has separated $r_{+}$ for different dimensions.
Figs. \ref{fig:1} (e) and (f) show variation of event horizon with $m$ for $d=4$ and $d=10$, respectively.
We can observe that for the fixed $l$ and $a$, $r_{+}$ varies with $m$.
It will be important to study behavior of temperature with $r_{+}$. Also, from the last equation of (\ref{B2}),
we should set $a^{2}<l^{2}$ to have positive entropy.\\
In the Fig. \ref{fig:2} we can see behavior of the temperature with $r_{+}$.
As  for the fixed $l$ and $a$, $r_{+}$ varies with $m$,
hence, $T$ varies with $r_{+}$. We can see different behavior for $d\geq6$ and $d\leq5$.
In the cases of $d=4$ and $d=5$ (solid and dashed lines of Fig. \ref{fig:2}),
temperature is totally increasing function of $r_{+}$.
On the other hand, for   $d\geq6$, temperature has a minimum. So,  it  decreases
with small $r_{+}$, and  increases with large $r_{+}$. Minimum temperature occurs at the critical value $r_{+c}$ ($d\geq6$),
which is root of the following equation,
\begin{equation}\label{B5}
\left( d-1 \right) {r_{+}}^{6}+ \left( 2\,{a}^{2}d-{l}^{2} \left( d-3
 \right)  \right) {r_{+}}^{4}+{a}^{2} \left(  \left( d-3 \right) {a}^{2}-2
\,{l}^{2} \left( d-6 \right)  \right) {r_{+}}^{2}-{a}^{4}{l}^{2} \left( d-
5 \right) =0.
\end{equation}
We can find critical value of event horizon radius by variation of $m$.
So, we observe that
$r_{+c}\approx1$ corresponding to $d=10$ and $d=4$ obtained for $m\approx1$ and $m\approx0.1$,  respectively (for $a=0.5$ and $l=1.3$).

Other thermodynamics quantities like  mass, angular momenta, angular velocity and volume can be expressed as
\begin{eqnarray}\label{B6}
M&=&\frac{m\omega_{d-2}}{4\pi\Xi^{2}}\left(1+\frac{(d-4)\Xi}{2}\right),
\\ \label{B7}
J&=&\frac{ma\omega_{d-2}}{4\pi\Xi^{2}},
\\ \label{B8}
\Omega&=&\frac{a}{l^{2}}\frac{r_{+}^{2}+l^{2}}{r_{+}^{2}+a^{2}},
\\ \label{B9}
V&=&\frac{Ar_{+}}{d-1}\left[1+\frac{a^{2}(r_{+}^{2}+l^{2})}{l^{2}\Xi(d-2)r_{+}^{2}}\right],
\end{eqnarray}
where $A=4s$, as given by the relation (\ref{B4}). It is clear that, in the special case of $d=4$ and $a=0$, we get  $V=\frac{4}{3}\pi r_{+}^{3}$ as expected.
Hence, the effect of $a$ is that it  increases  the  volume in four dimension. Situation is similar for the higher dimensional case. Finally, Gibbs free energy given by
\begin{equation}\label{B10}
G=\frac{\omega_{d-2}r_{+}^{d-5}}{16\pi\Xi^{2}}\left(3a^{2}+r_{+}^{2}-\frac{(r_{+}^{2}-a^{2})^{2}}{l^{2}}
+\frac{3a^{2}r_{+}^{4}+a^{4}r_{+}^{2}}{l^{4}}\right).
\end{equation}
Behavior of $G$ and critical points discussed.
We will discuss  other thermodynamics potentials like Helmholtz free energy, $PV$ diagram,
critical point and stability of system in the next section, and also analyze the  effect of  thermal fluctuations on the thermodynamics of this system.

\section{Thermal Fluctuations}
\begin{figure}[h!]
 \begin{center}$
 \begin{array}{cccc}
\includegraphics[width=50 mm]{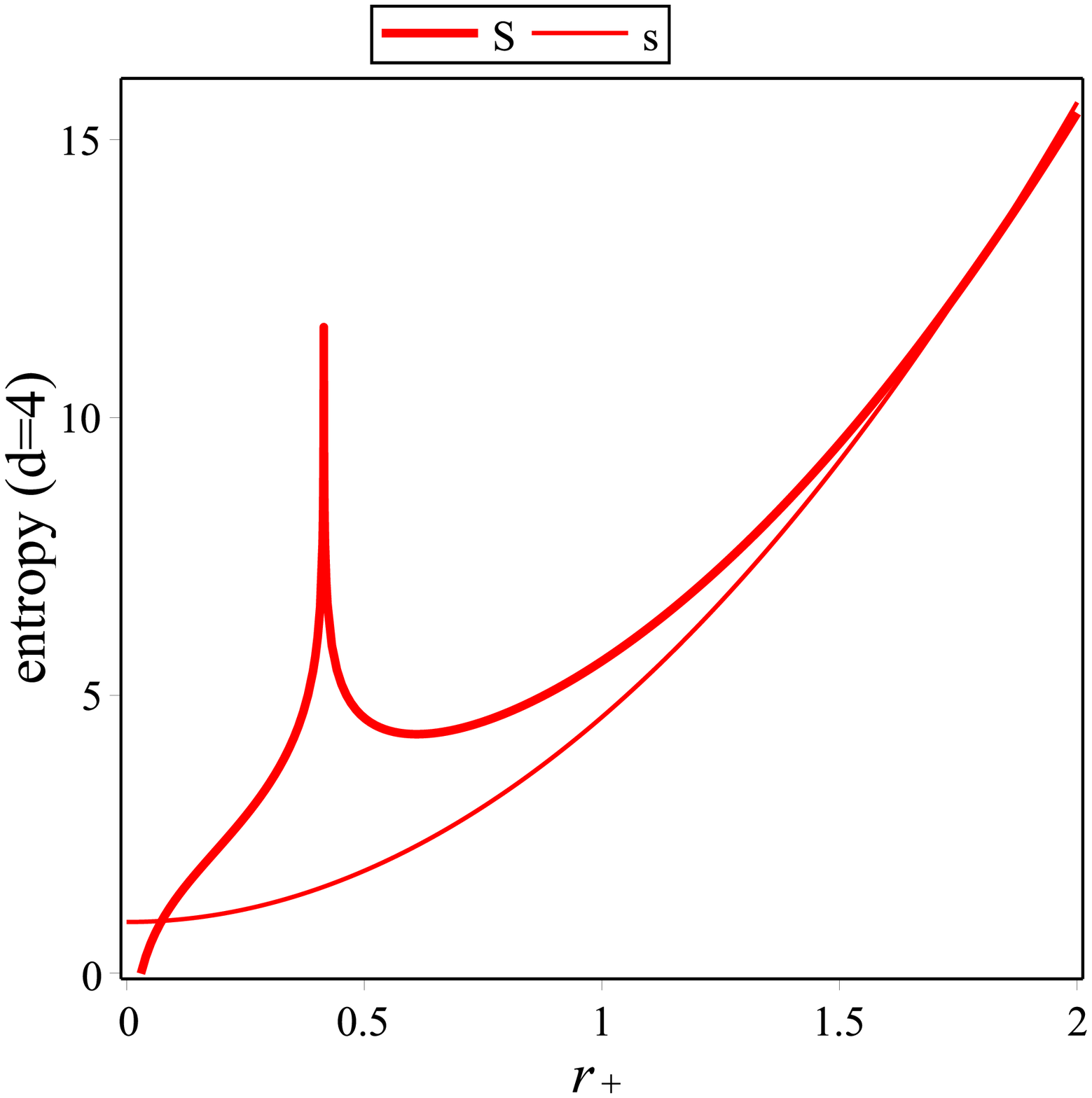}&\includegraphics[width=50 mm]{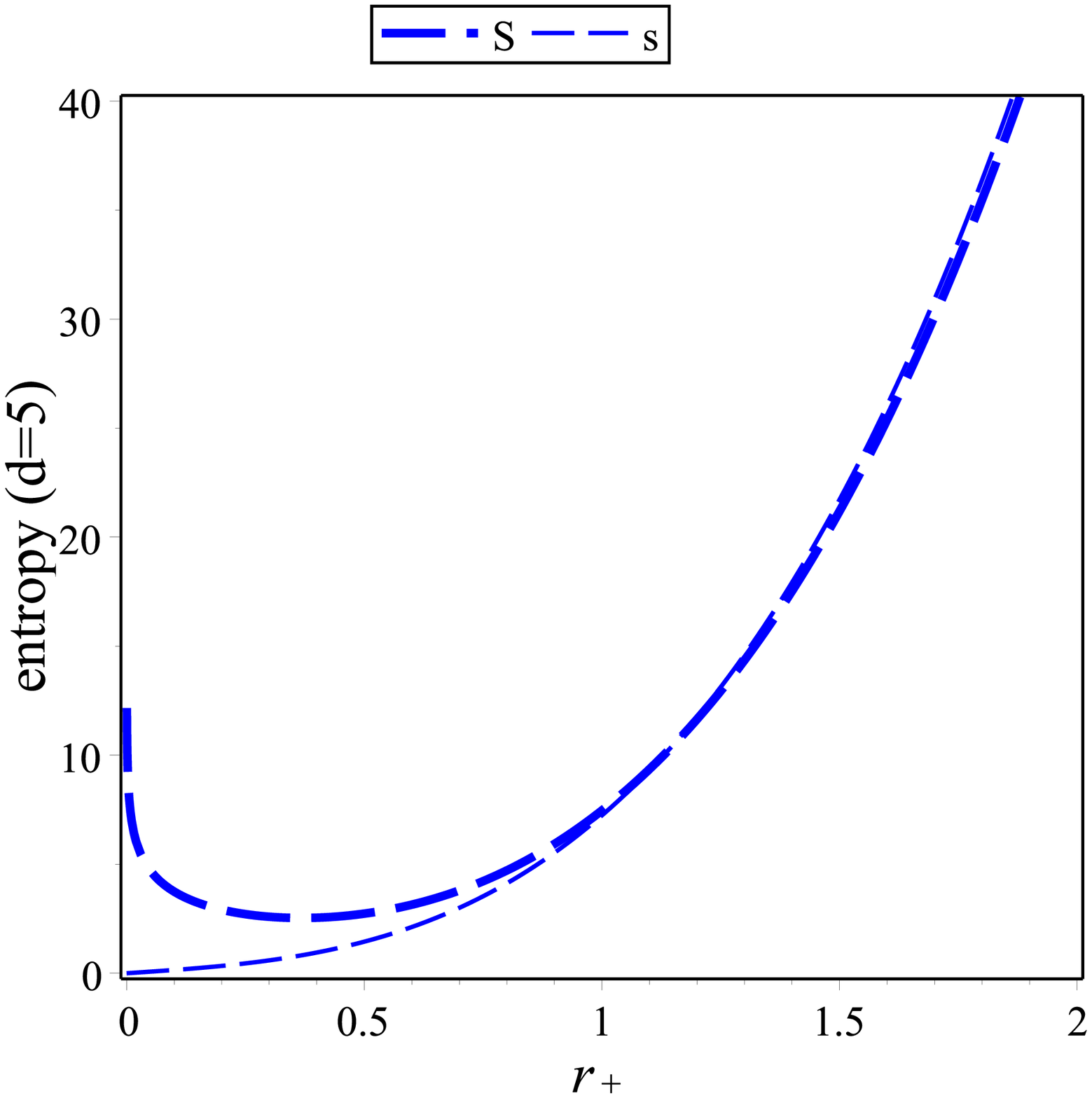}\\
\includegraphics[width=50 mm]{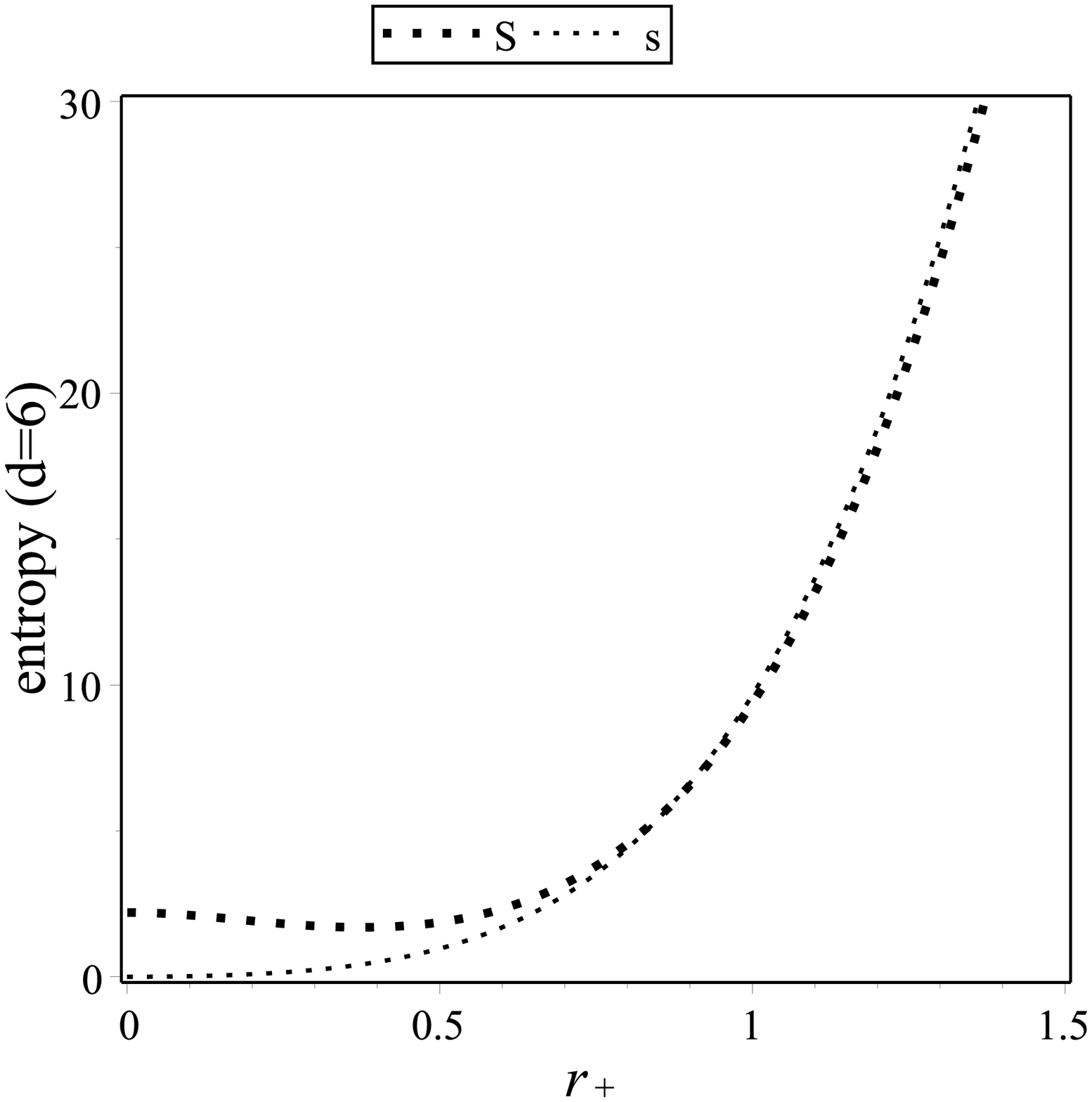}&\includegraphics[width=50 mm]{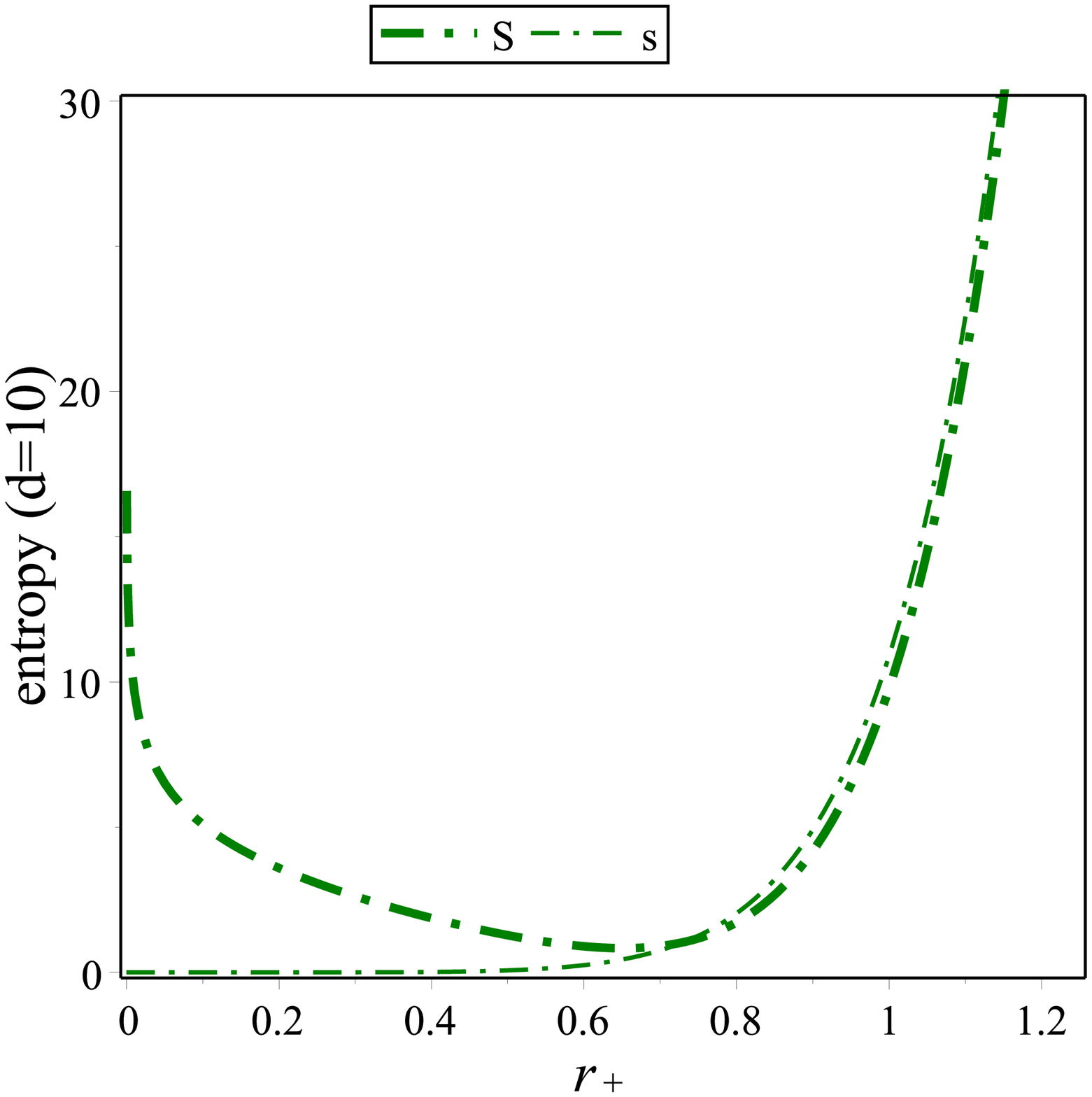}
 \end{array}$
 \end{center}
\caption{Entropy in terms of $r_{+}$ for $l=1.3$, and $a=0.5$. Thin curves show ordinary entropy ($\alpha=0$) given by the equation (\ref{B4}),
while thick curves show logarithmic corrected entropy ($\alpha=1$) given by the equation (\ref{C2}). $d=4$ (solid red), $d=5$ (dashed blue), $d=6$
(dotted black), $d=10$ (dash dotted green).}
 \label{fig:3}
\end{figure}

The entropy of a black hole will be corrected by a logarithmic term due to the thermal fluctuations.
The
It may be noted that we will be analyzing this system very close to the equilibrium, and so we will analyze
the thermal fluctuations as perturbations around the equilibrium. This approximation will be valid as long
as the correction due to the    thermal fluctuations is small compared to the original quantity, i.e.,
as long as $S- s/ s << 1$, where $S$ is the corrected entropy and $s$ is the original entropy of the system.
It should also be noted that at very high temperatures, i.e., near the Planck scale,
the thermal fluctuations will be very large, and at this stage the system cannot be analyzed as a perturbation around equilibrium temperature.
However, for such temperatures, where we can analyze this system as a perturbation around equilibrium, we can write
the corrected entropy  as  \cite{Landau, fl, l1},
\begin{equation}\label{C1}
S = s - \frac{\ln  s^{\prime\prime}}{2},
\end{equation}
here prime denote derivative with respect to $ {T}^{-1}$, where $T$ is the equilibrium temperature.
Furthermore, for such systems,
 the   second derivative of the entropy can be expressed in  terms of  fluctuations of the energy, and so the
  the   corrected entropy can be written as \cite{l1}
\begin{equation}\label{C2}
S = s -\frac{\alpha}{2} \ln(sT^{2}),
\end{equation}
where $s$ is original entropy given by the Eq. (\ref{B4}). The parameter $\alpha$ is added by hand to analyze effect of thermal fluctuations.
Furthermore,
as the logarithmic corrections to the entropy are generated from almost all approaches to quantum gravity, but the coefficient of such
corrections varies between different approaches, we will keep such a coefficient as a variable in this paper. Thus, we can effectively
discuss the effect of the corrections to the thermodynamics from various different approaches to quantum gravity. It may be noted that   the limiting cases,
 $\alpha=1$ is valid for the  the entropy corrected by thermal fluctuations, and    $\alpha=0$ is valid for the original entropy of this system.
\\
As we will be analyzing a very general form of the corrections to the entropy, we will generalize this result.
We can generalize this result by looking at the ideal gas entropy,
written as,
\begin{equation}\label{IGS}
S = \frac{5}{2}N k_{B} -N k_{B}\ln(\frac{N}{V}\frac{h^{3}}{(2\pi m k_{B} T)^{\frac{3}{2}}}),
\end{equation}
where $k_{B}$ is Boltzmann constant and $N$ is the total particle number. Motivated by the equation (\ref{IGS}), we propose the following logarithmic entropy,
\begin{equation}\label{LCS}
S(T) = s +\alpha\ln(s f(T)),
\end{equation}
where $f(T)$ is specific function of temperature,  and $\alpha$ is free parameter  which is used to incorporate  the
the effect of logarithmic corrections.
If we assume $s=\frac{5}{2}N k_{B}$, $\alpha=-N k_{B}$ and $f(T)=\frac{2\hbar}{5V}(\frac{2\pi}{m})^{\frac{3}{2}}T^{-\frac{3}{2}}$, then
the ideal gas entropy (\ref{IGS}) reproduced. Presence of $\hbar$ in the $f(T)$ show that thermal fluctuations are indeed quantum effect
and will be tested for several physical system.\\
Furthermore, if we assume $f(T)=T^{2}$ and $\alpha=-\frac{1}{2}$, then the corrections to the  entropy due to thermal fluctuations are   reproduced.\\
In the canonical ensemble, one can relate entropy $S(T)$ to the partition function $Z$,
\begin{equation}\label{P1}
S(T) = k_{B}\ln{Z}+k_{B} T (\frac{\partial \ln{Z}}{\partial T}),
\end{equation}
So, for the any physical system with ordinary entropy $s$, one can obtain,
\begin{equation}\label{P2}
\ln{Z}=\frac{1}{T}\int{\frac{s +\alpha\ln(s f(T))}{k_{B}}dT}.
\end{equation}
Using the equation (\ref{P2}), one can obtain other thermodynamics quantities like internal energy,
\begin{equation}\label{P3}
E=k_{B}T^{2}\frac{d\ln{Z}}{dT}=T(s +\alpha\ln(s f(T)))-\int{(s +\alpha\ln(s f(T)))dT},
\end{equation}
Helmholtz free energy,
\begin{equation}\label{P4}
F=-k_{B}T\ln{Z}=-\int{(s +\alpha\ln(s f(T)))dT},
\end{equation}
specific heat at constant volume,
\begin{equation}\label{P5}
C_{V}=T\left[\frac{d}{d T}(s +\alpha\ln(s f(T)))\right]_{V},
\end{equation}
specific heat at constant pressure,
\begin{equation}\label{P6}
C_{P}=C_{V}+\left[P+\left(\frac{\partial E}{\partial V}\right)_{T}\right]\left(\frac{\partial V}{\partial T}\right)_{P},
\end{equation}
and  pressure,
\begin{equation}\label{P7}
P=k_{B}T\left(\frac{d\ln{Z}}{dV}\right)_{T}.
\end{equation}
Then, we can obtain enthalpy and Gibbs free energy as
\begin{equation}\label{P8}
H=E+PV,
\end{equation}
and
\begin{equation}\label{P9}
G=F+PV=H-T(s +\alpha\ln(s f(T))).
\end{equation}
Now we will assume $f(T)=T^{2}$,  and investigate thermodynamics properties of this system.\\
In the plots of the Fig. \ref{fig:3}, we can see the behavior of the entropy for various dimensions (we focus on $d=4, 5, 6, 10$).
It is expected that logarithmic correction is important for small black hole, so we see that for the larger $r_{+}$ both
$S$ and $s$ coincide. On the other hand, for the small $r_{+}$, corrected entropy has completely different behavior,
so a minimum entropy is available for $d\geq5$. In the special case of $d=4$, there is a maximum for the entropy, which has a   singular behavior.
However, it is possible that this approximation breaks down near this point, as the perturbations can not be used to analyze such
points.

\begin{figure}[h!]
 \begin{center}$
 \begin{array}{cccc}
\includegraphics[width=50 mm]{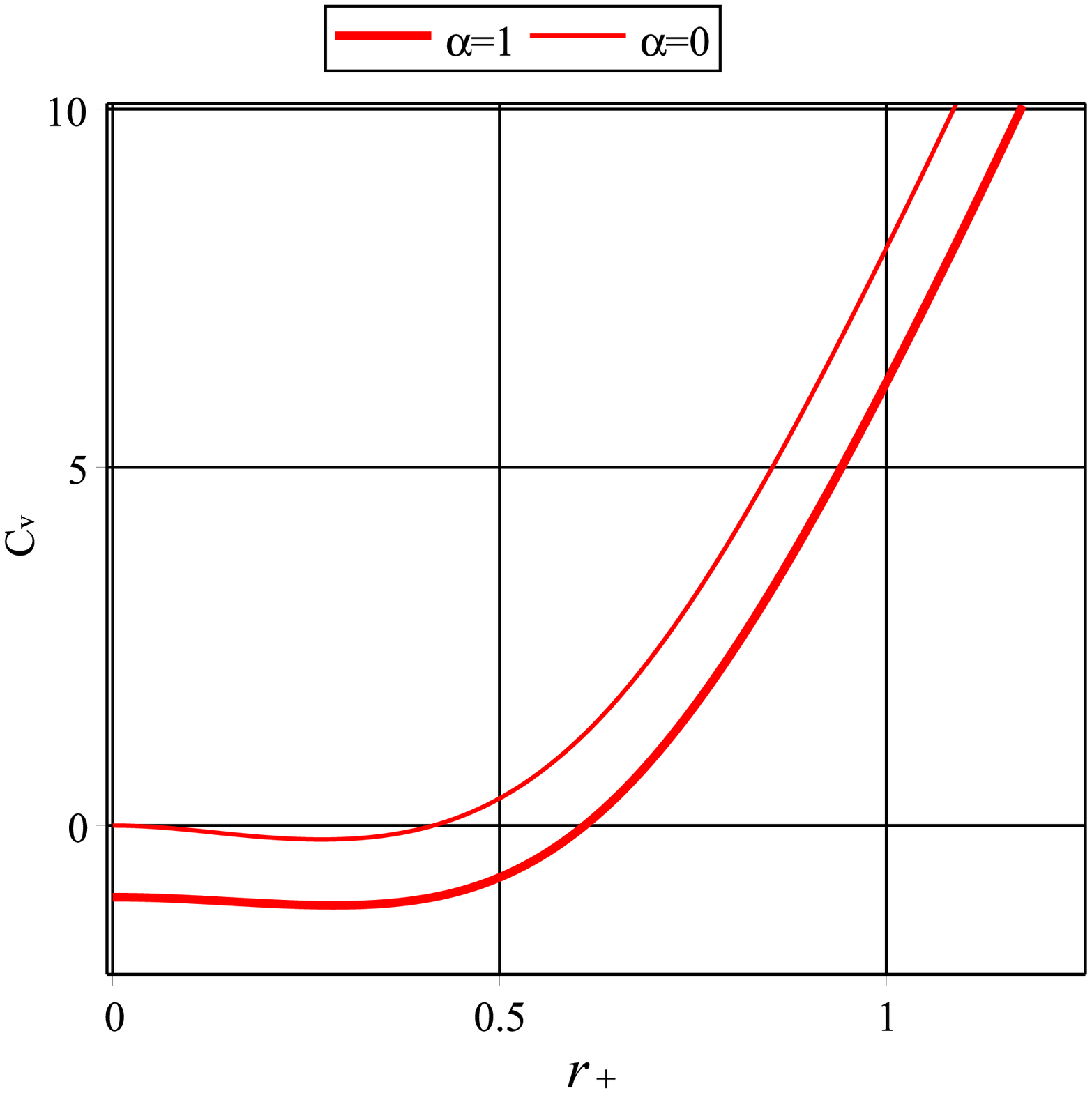}&\includegraphics[width=50 mm]{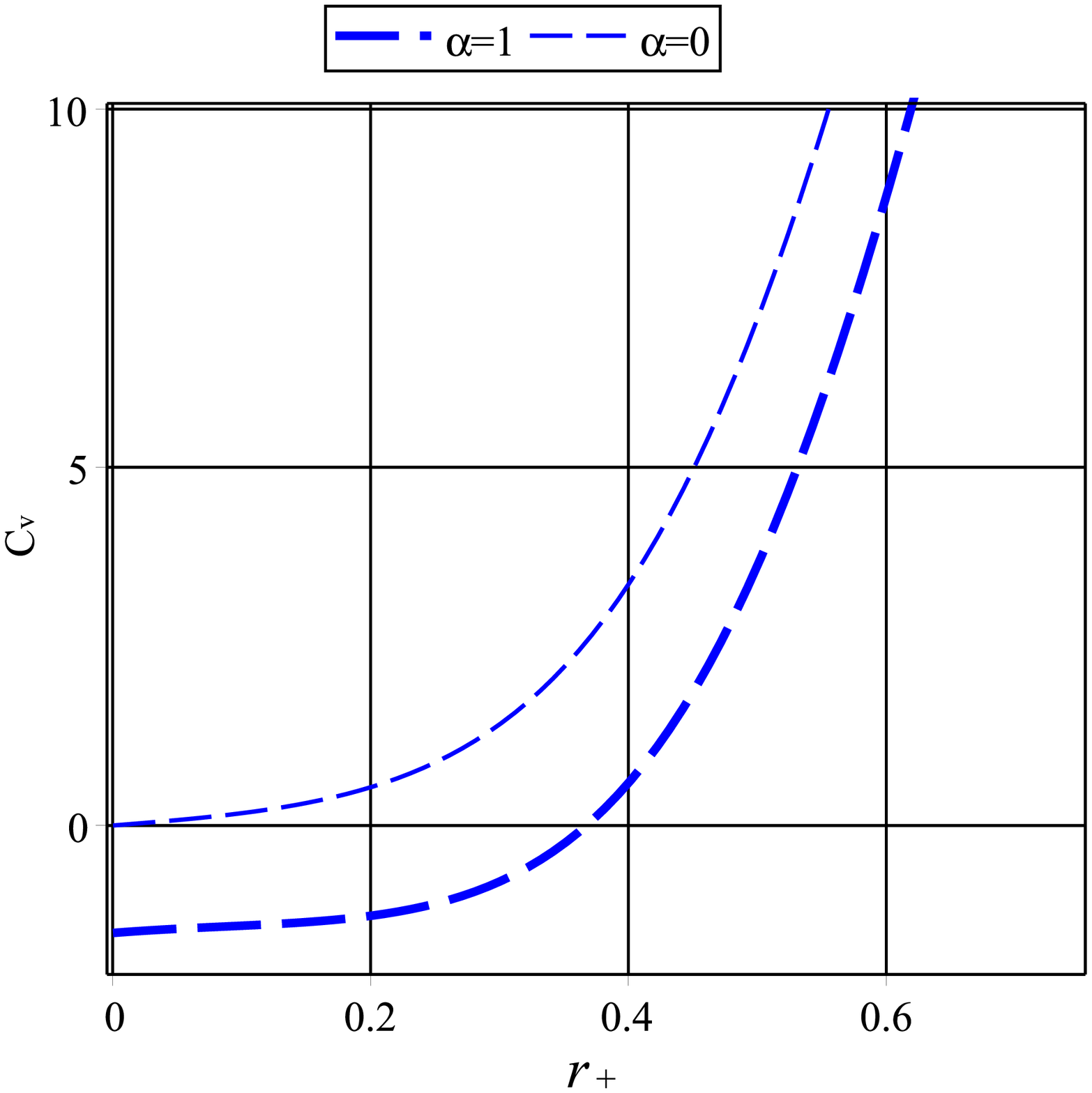}\\
\includegraphics[width=50 mm]{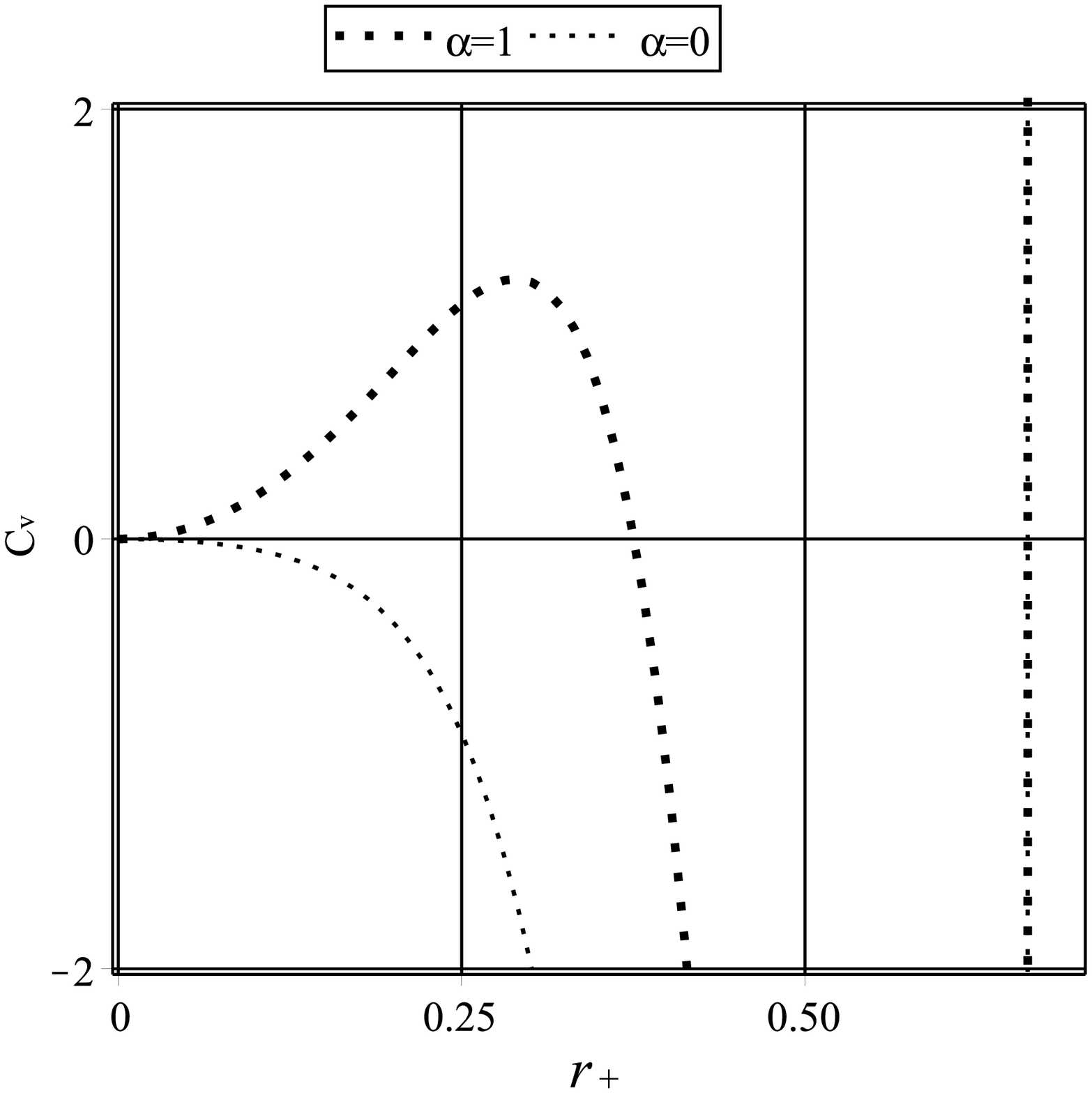}&\includegraphics[width=50 mm]{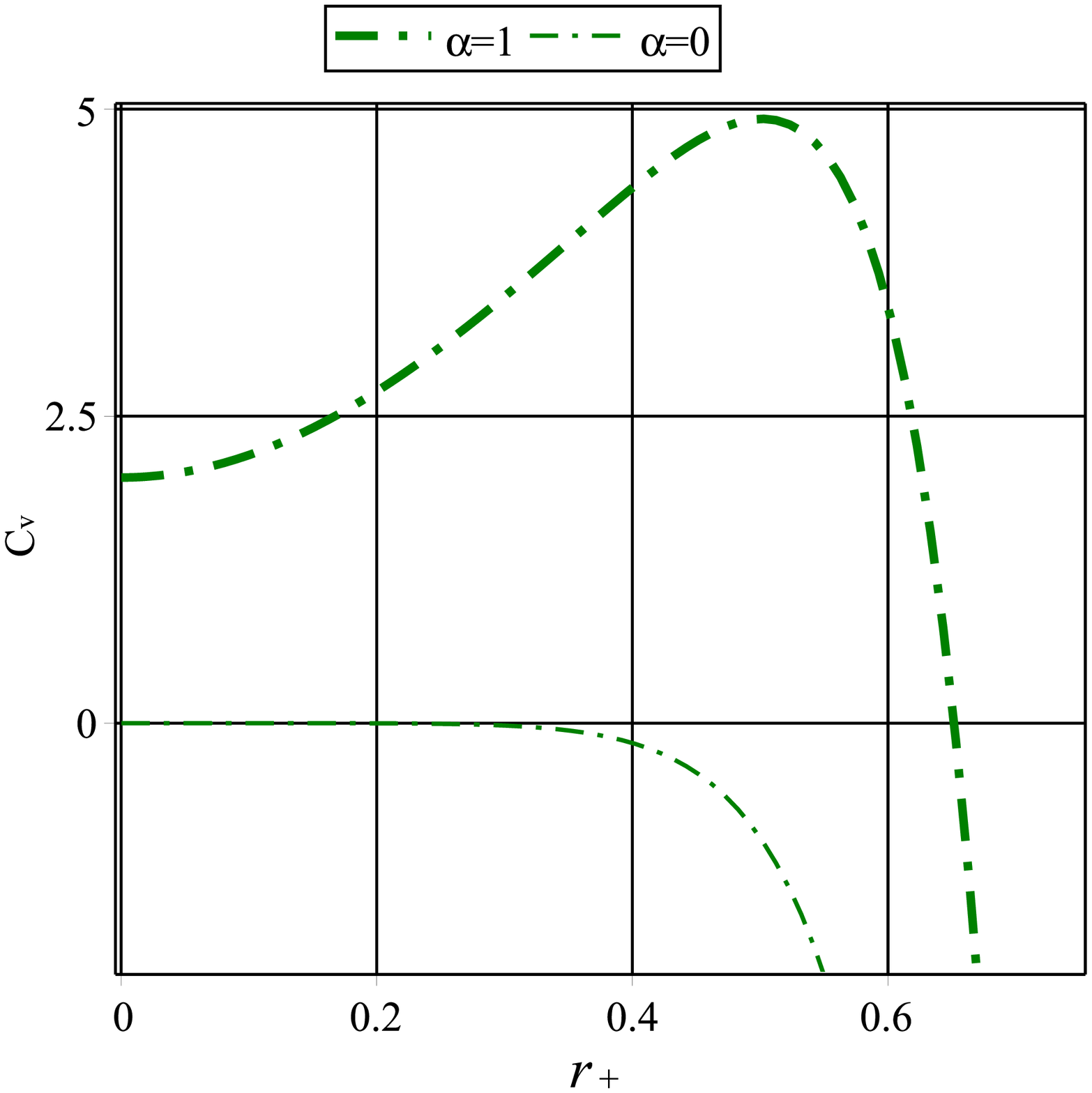}
 \end{array}$
 \end{center}
\caption{Specific heat in terms of $r_{+}$ for $l=1.3$, and $a=0.5$. Thin curves show ordinary case ($\alpha=0$), while thick curves show logarithmic
corrected case ($\alpha=1$). $d=4$ (solid red), $d=5$ (dashed blue), $d=6$ (dotted black), $d=10$ (dash dotted green).}
 \label{fig:4}
\end{figure}

Now, we can use the specific heat given by the equation (\ref{P5})
to study stability of the black hole. Plots of the Fig. \ref{fig:4} show behavior of
specific heat in terms of $r_{+}$ for various dimensions. In absence of extra dimensions ($d=4$),
we can see  unstable region for the small black hole. It means that there is a minimum size
for the black hole. So, the  thermal fluctuations can cause a  phase transition in the black hole,
 (as one can see from the second, third and last plots of
Fig. \ref{fig:4}), and so  without the logarithmic corrections there is no phase transition, except in four dimensions.
We can see that the black hole is completely stable in $d=5$, without logarithmic correction.
However, thermal fluctuations make the black hole unstable. Situation is completely different for the cases of $d\geq6$.
In this cases the black hole is completely unstable.
It is an interesting result that the
logarithmic corrections are needed to make the black holes stable  in presence of at least two extra dimensions ($d\geq6$).

\begin{figure}[h!]
 \begin{center}$
 \begin{array}{cccc}
\includegraphics[width=50 mm]{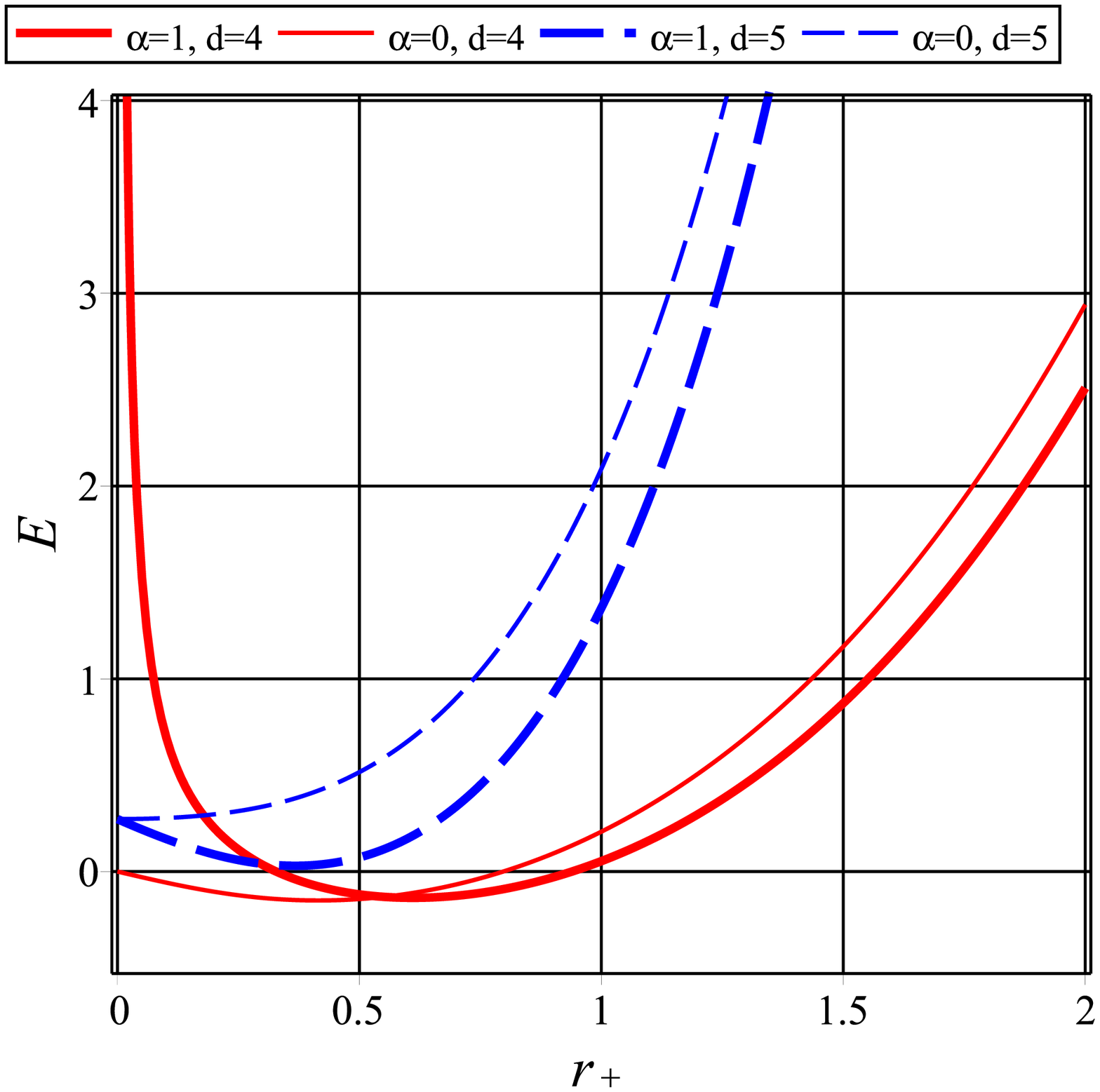}&\includegraphics[width=50 mm]{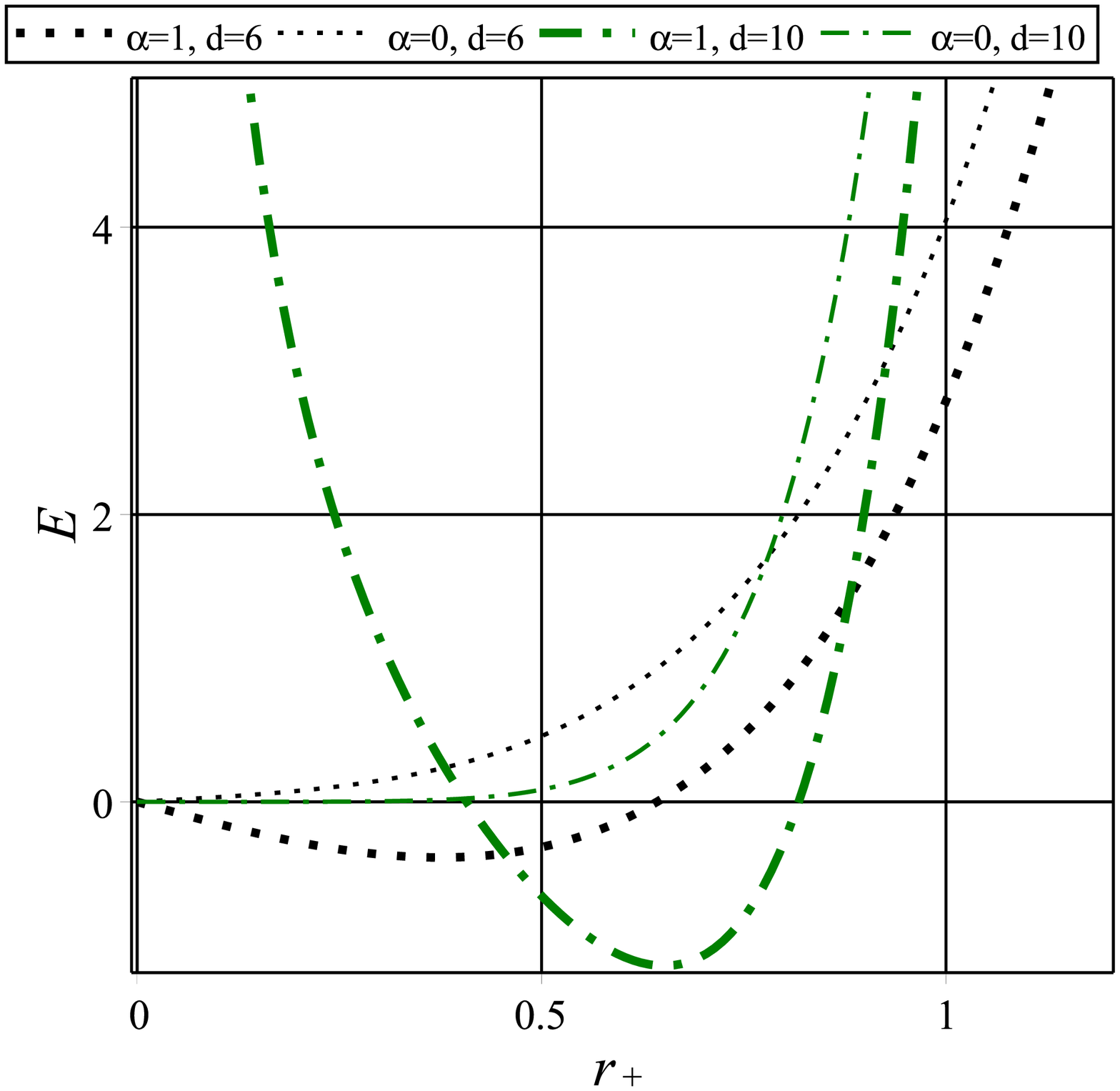}
 \end{array}$
 \end{center}
\caption{Internal energy in terms of $r_{+}$ for $l=1.3$, and $a=0.5$. Thin curves show ordinary case ($\alpha=0$),
while thick curves show logarithmic corrected case ($\alpha=1$). $d=4$ (solid red), $d=5$ (dashed blue), $d=6$ (dotted black), $d=10$ (dash dotted green).}
 \label{fig:5}
\end{figure}

In order to obtain Helmholtz free energy we need to calculate internal energy,
\begin{equation}\label{C4}
E=\int{C dT}=E_{1}+\alpha E_{2},
\end{equation}
where $E_{1}$ and $E_{2}$ for the case of $d=4$ given by,
\begin{equation}\label{C5}
E_{1}=\frac{2a(l^{2}-a^{2})\tan^{-1}(\frac{r_{+}}{a})+r_{+}(2a^{2}-l^{2}-r_{+}^{2})}{2(a^{2}-l^{2})},
\end{equation}
and
\begin{equation}\label{C6}
E_{2}=\frac{ar_{+}(a^{2}+r_{+}^{2})\tan^{-1}(\frac{r_{+}}{a})+\frac{a^{2}}{4}(l^{2}-5r_{+}^{2})-\frac{3}{2}r_{+}^{4}}{\pi l^{2}r_{+}(a^{2}+r_{+}^{2})}.
\end{equation}
In presence of extra dimension, we can obtain approximately similar results which are illustrated by Fig. \ref{fig:5}.
In the case of $d=4$, we can see important effect for the small black hole, which have  infinite energy because of
 thermal fluctuations. In the cases of $d=6$ and $d=10$, we obtain  a minimum for internal energy, due to the thermal fluctuations.

\begin{figure}[h!]
 \begin{center}$
 \begin{array}{cccc}
\includegraphics[width=50 mm]{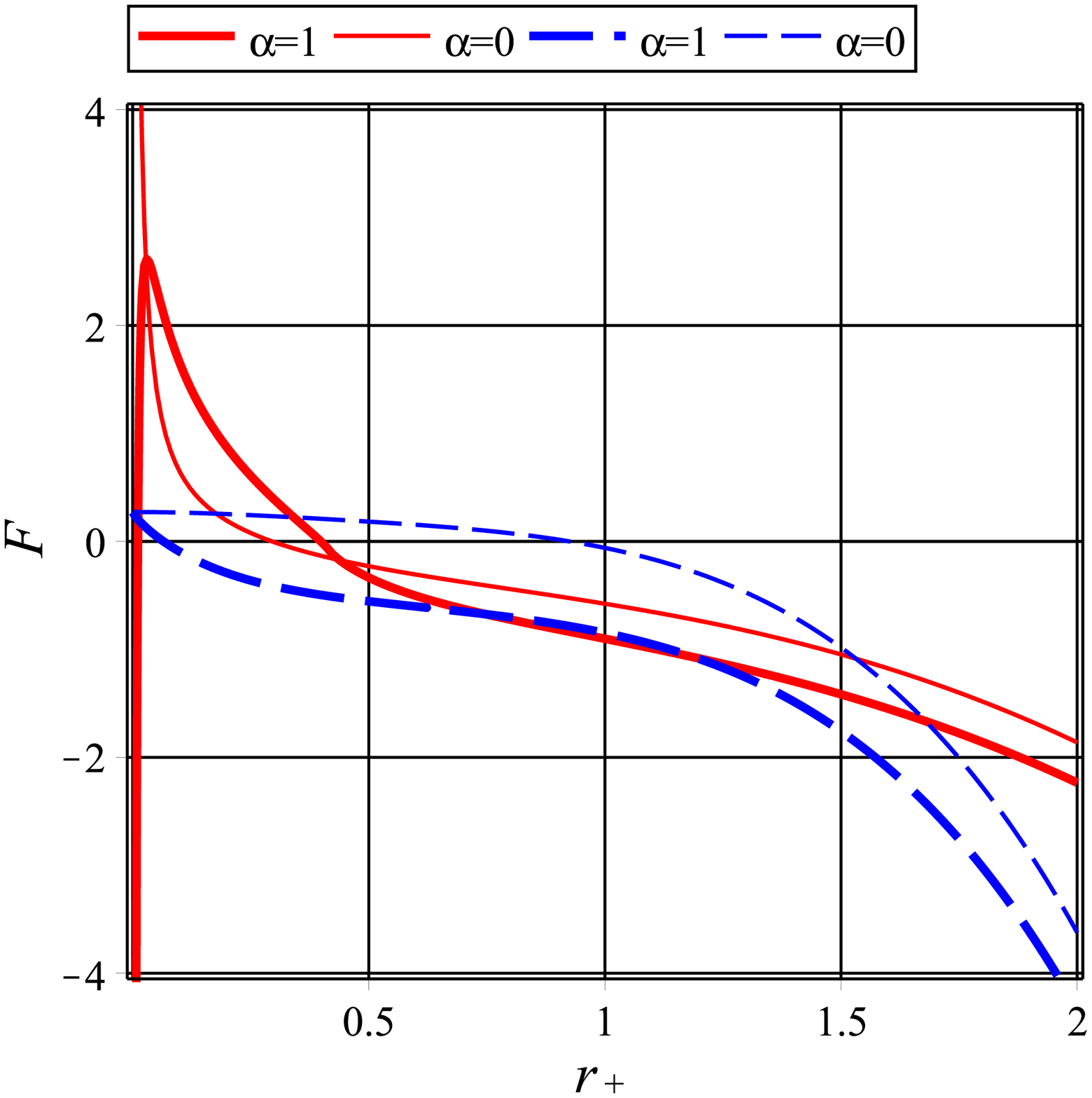}&\includegraphics[width=50 mm]{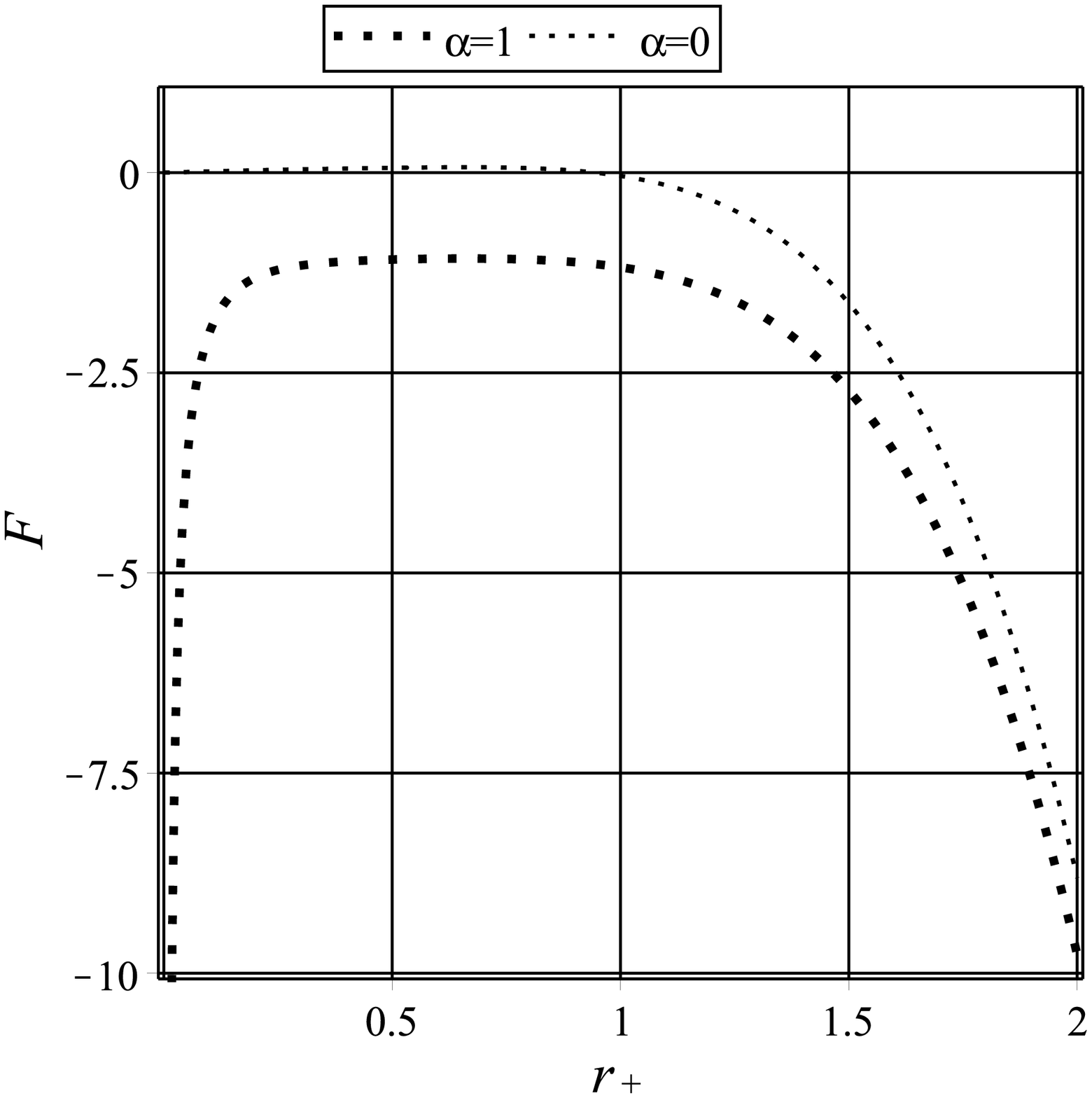}&\includegraphics[width=50 mm]{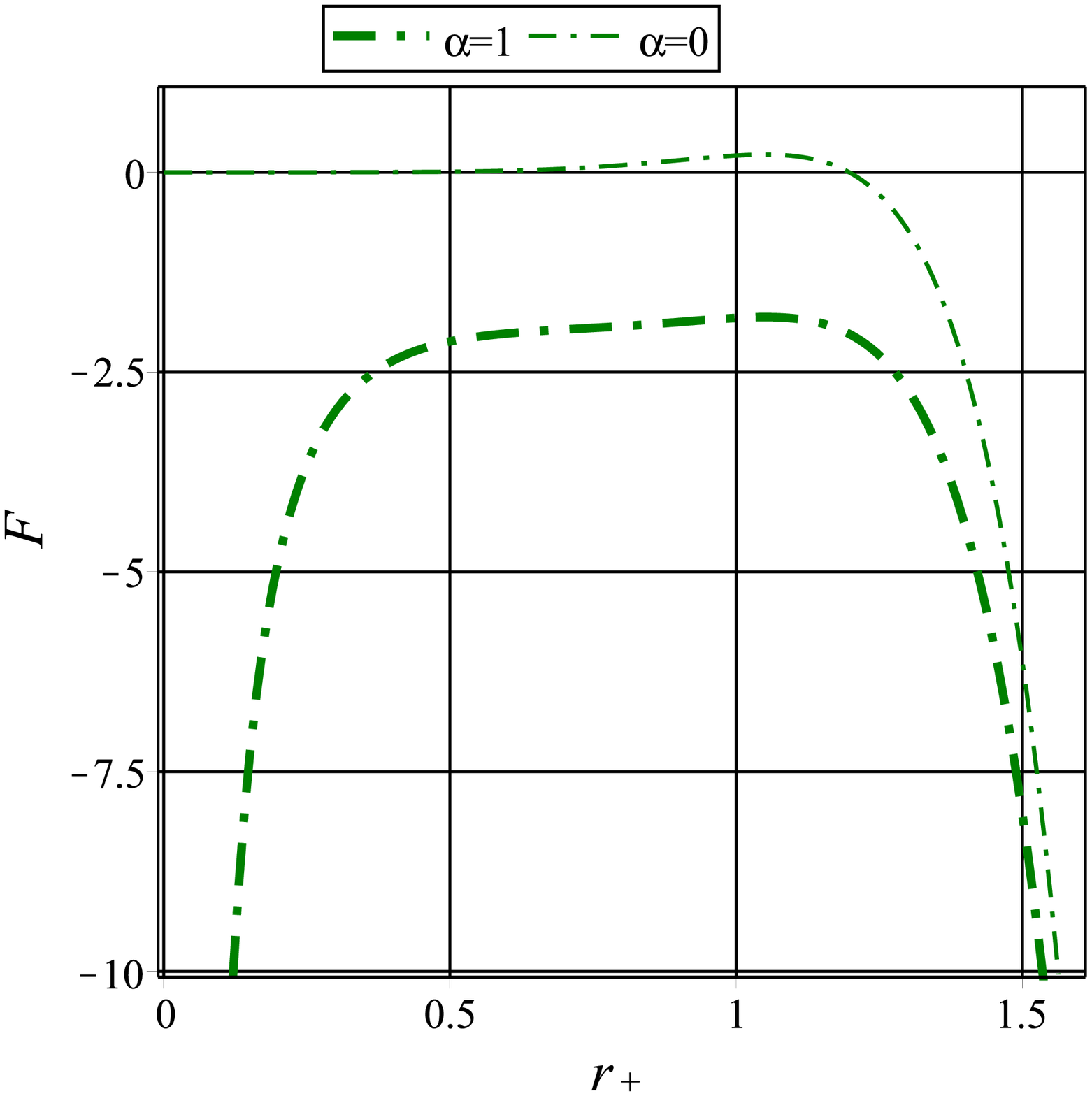}
 \end{array}$
 \end{center}
\caption{Helmholtz free energy in terms of $r_{+}$ for $l=1.3$, and $a=0.5$. Thin curves show ordinary case ($\alpha=0$),
while thick curves show logarithmic corrected case ($\alpha=1$). $d=4$ (solid red), $d=5$ (dashed blue), $d=6$ (dotted black), $d=10$ (dash dotted green).}
 \label{fig:6}
\end{figure}

Now, we can calculate Helmholtz free energy given by the equation (\ref{P4}).
In the plots of the Fig. \ref{fig:6}, we can see behavior of the Helmholtz
free energy in terms of $r_{+}$ for various dimensions.
In the cases of $d\geq5$, we can see that logarithmic correction decreases value of $F$.
In the special case of $d=4$, there is a critical horizon radius $r_{c}$, where $F(\alpha=1)=F(\alpha=0)$.
If $r_{+}>r_{c}$, then the effect of logarithmic correction is that it  decreases the  Helmholtz free energy,
while if $r_{+}<r_{c}$, then the effect of logarithmic correction is that it increases the  Helmholtz free energy to a maximum value.
Now,  $F(\alpha=0)\rightarrow+\infty$,  and  $F(\alpha=1)\rightarrow-\infty$
at $r_{+}\rightarrow0$. Interesting point is that value of corrected
$F$ at $r_{+}\approx1$ is the same for   $d=4$ and $d=5$.
In higher dimensions, Helmholtz free energy is zero for small black hole without thermal fluctuations,
while it has negative infinite value in presence of thermal fluctuations.

Gibbs potential obtained using enthalpy (which interpreted as mass given by the equation (\ref{B6})), temperature and entropy which is given by the equation (\ref{P9}).
We can see that value of $G$ is depend on $m$, $a$ and $l$. So, at the fixed $m$, $a$ and $l$, value of $r_{+}$ is fixed.
In the table (1),  we can see effect of logarithmic correction on the Gibbs potential. We can conclude that
thermal fluctuations decrease value of $G$ in the $d=4$ and $d=5$ dimensions. On the
other hand, for the cases of $d\geq6$, thermal fluctuation increases value of the Gibbs potential.
For the cases of $d=9$ and $d=10$, Gibbs free energy is negative in absence of logarithmic correction, but
it is positive in presence of thermal fluctuations.

\begin{table}[h!]
  \centering
  \caption{Value of the Gibbs potential for $l=1.3$, $m=1$ and $a=0.5$ for various dimensions.}
  \label{tab:table1}
  \begin{tabular}{|l|c||r|}
    \hline
    $\alpha=0$          & $\alpha=1$   & $d$  \\ \hline\hline
    0.38932             & 0.21797      & 4     \\ \hline
    0.47997             & 0.4132       & 5     \\ \hline
    0.46452             & 0.5807       & 6     \\ \hline
    0.34055             & 0.6809       & 7     \\ \hline
    0.14279             & 0.7235       & 8     \\ \hline
    -0.076259           & 0.7434       & 9   \\  \hline
    -0.26761            & 0.7751       & 10    \\  \hline
  \end{tabular}
\end{table}

\begin{figure}[h!]
 \begin{center}$
 \begin{array}{cccc}
\includegraphics[width=50 mm]{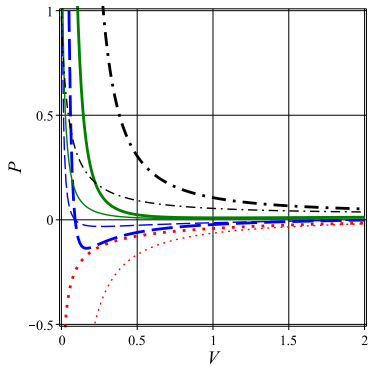}
 \end{array}$
 \end{center}
\caption{Pressure in terms of $V$ for $l=1.3$ and $d=5$. Thin curves show ordinary case ($\alpha=0$), while thick
curves show logarithmic corrected case ($\alpha=1$). $a=0$ (dotted red), $a=0.25$ (dashed black), $a=0.333$ (solid green), $a=0.5$ (dash dotted black).}
 \label{fig:7}
\end{figure}

Now, using the equations (\ref{B9}), (\ref{P4}) and
\begin{equation}\label{C9}
P=-\left(\frac{\partial F}{\partial V}\right),
\end{equation}
we can study behavior of pressure. In order to find $PV$ diagram, we can find $r_{+}$ in terms of $V$ from the Eq.
(\ref{B9}), then remove $r_{+}$ in the Eq.  (\ref{C9}), to obtain behavior of $P$ in terms of $V$. In the   case of $d=5$, we can obtain,
\begin{equation}\label{C10}
r_{+}^{2}=\frac{\pi l a^{2}(2l^{2}-a^{2})-(a^{2}-l^{2})\sqrt{\pi^{2} a^{4} l^{2}+6(3l^{2}-2a^{2})V}}{\pi l (2a^{2}-3l^{2})}.
\end{equation}
Using the relation (\ref{C10}) in the Eq.  (\ref{C9}) for $d=5$,  we can obtain behavior of $P$
as illustrated in the Fig. \ref{fig:7}. We can see critical point shifted
due to the thermal fluctuations. It is clear that the interested case  is when
$l=1.3$ and, $a=0.5$, and for this we obtain a completely stable black hole. We should note that, there is some negative regions in the Fig. \ref{fig:7}
which are unphysical since it would require analytically continuing $l\rightarrow il$,  and so we cannot use the values  below $P=0$ to analyze the
  Van der waals behavior. These unphysical situations are obtained for small $a$ and large $l$, so one can remove such negative region
by fixing $a$ and $l$. It produces    a maximum value for $l$ and minimum value for $a$, with an uncertainty like equation.
 Fitting the curves of Fig. \ref{fig:7},  suggests the following Virial expansion form of pressure,
\begin{equation}\label{C11}
P=\frac{A}{V}+\frac{B}{V^{2}}+\frac{C}{V^{3}}+\cdots
\end{equation}
and hence we can write,
\begin{equation}\label{C12}
\frac{PV}{T}=A(T)+\frac{B(T)}{V}+\frac{C(T)}{V^{2}}+\cdots
\end{equation}
It means that singly spinning Kerr-AdS black hole in five dimension behave as van der Waals fluid. The best fitted values of a stable case for Virial
coefficients $A(T)$, $B(T)$ and $C(T)$ yields to the following relation,
\begin{equation}\label{C12}
\frac{PV}{T}=0.3+\frac{0.7}{V^{2}},
\end{equation}
  so  $B(T)=0$, and a black hole is in Boyle temperature ($B(T)=0$).

\begin{figure}[h!]
 \begin{center}$
 \begin{array}{cccc}
\includegraphics[width=50 mm]{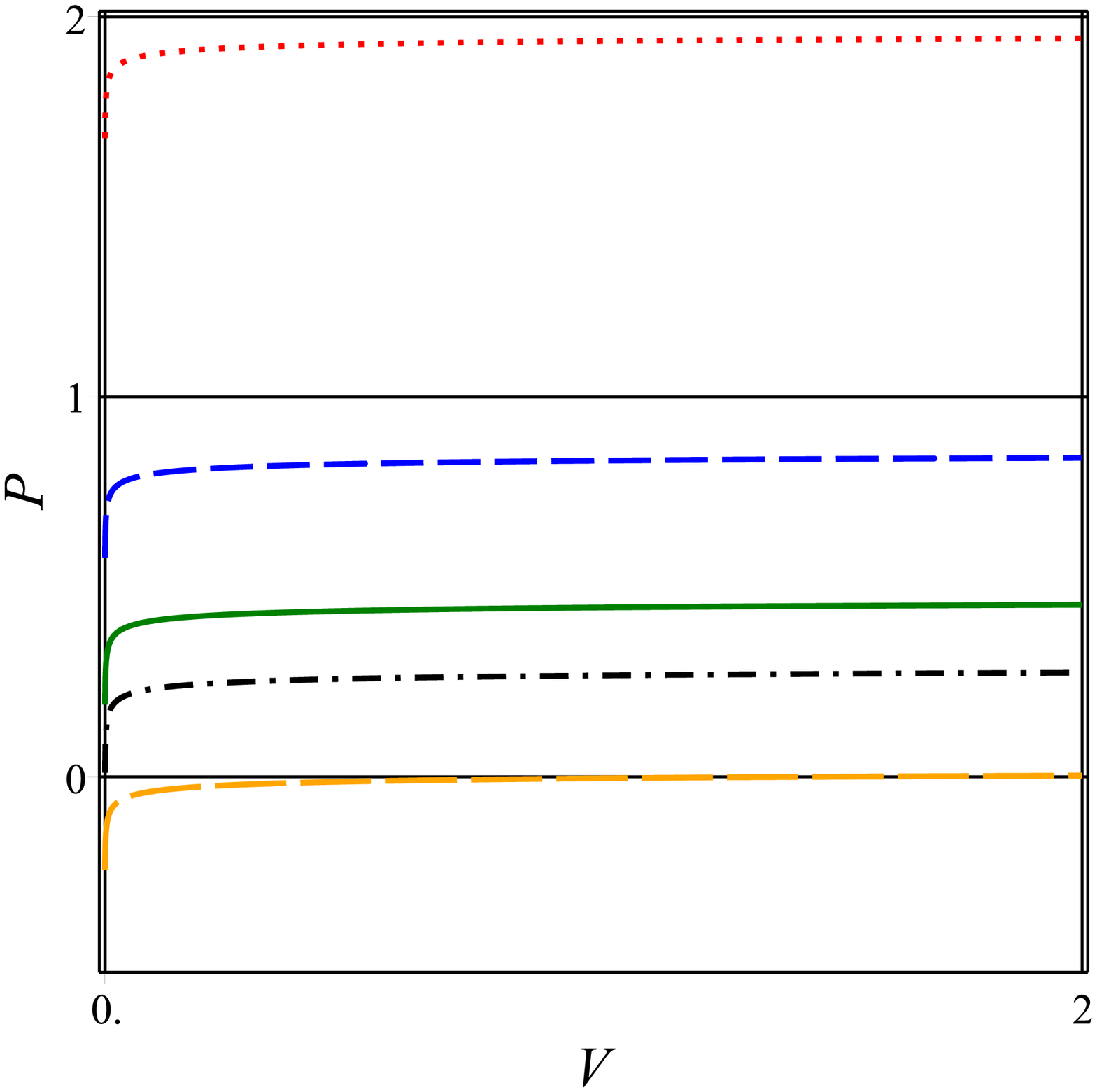}&\includegraphics[width=50 mm]{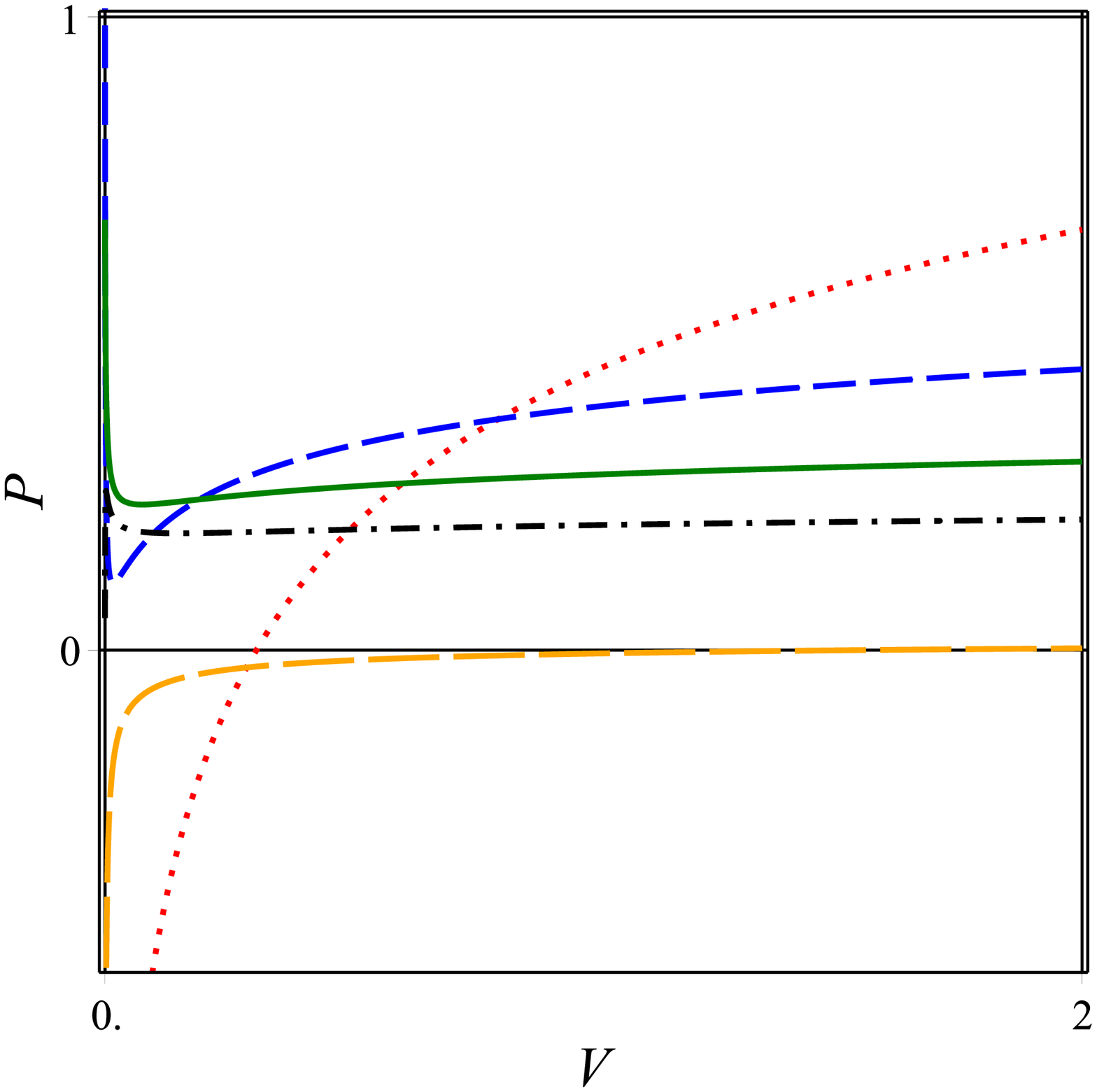}
 \end{array}$
 \end{center}
\caption{Pressure in terms of $V$ for $a=0.01$ and $d=10$.
Left plot show ordinary case ($\alpha=0$), while right plot show logarithmic corrected case ($\alpha=1$).
$l=0.2$ (dotted red), $l=0.3$ (dashed blue), $l=0.4$ (solid green), $l=0.5$ (dash dotted black), $l=1.3$ (long dashed orange).}
 \label{fig:8}
\end{figure}

Situation is complicated for other dimensions where analytic expression of $r_{+}$ in terms of $V$ is not available.
For example, using the relation (\ref{B9}), with $d=10$, we have the following equation,
\begin{equation}\label{C13}
(8l^{2}-7a^{2})r^{9}+(9a^{2}l^{2}-7a^{4})r^{7}+l^{2}a^{4}r^{5}-\frac{945(a^{2}-l^{2})^{2}}{l^{2}\pi^{4}}V=0.
\end{equation}
However, we can assume small $a$ and do numerical analysis to obtain pressure in terms of volume, and this is    illustrated by the Fig. \ref{fig:8}.
We can see that the thermal fluctuations are necessary to have critical point, and
obtain expected curves. Right plot of the Fig. \ref{fig:8}, shows that $a$ and $l$ should be small to have the expected curves.

\section{Four Dimensional Kerr-AdS Black Hole}
\begin{figure}[h!]
 \begin{center}$
 \begin{array}{cccc}
\includegraphics[width=50 mm]{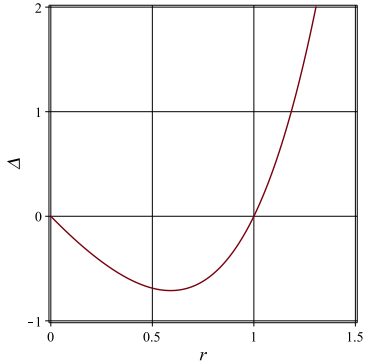}
 \end{array}$
 \end{center}
\caption{Typical behavior of $\Delta$ in terms of $r$ with $l=1$, $m=1$, and $a=0$.}
 \label{fig:9}
\end{figure}

In this section, we will discuss about the special case of a rotating black hole in four dimensions, and also analyze the effects of thermal fluctuation
on such a black hole.
The corresponding metric is obtained by putting $d=4$ in the solution given by (\ref{B1}) and (\ref{B2}).
Thermodynamics quantities of Kerr-AdS black hole are listed below. The black hole mass and angular momenta are given by
\begin{equation}\label{A33}
M=\frac{m}{\Xi^{2}},
\end{equation}
and
\begin{equation}\label{A44}
J=\frac{ma}{\Xi^{2}},
\end{equation}
while angular velocity of the horizon is given by
\begin{equation}\label{A55}
\Omega=\frac{a(r_{+}^{2}+l^{2})}{l^{2}(r_{+}^{2}+a^{2})},
\end{equation}
where $r_{+}$ is the largest root of $\Delta = 0$. By appropriate choice of $l$, $m$ and $a$, there are two real positive roots,
$r_{\pm}$, which illustrated by the Fig. \ref{fig:9}. In the table (2), we can see value of $r_{+}$, with some possible values of free black hole parameters.

\begin{table}[h!]
  \centering
  \caption{Value of $r_{+}$.}
  \label{tab:table2}
  \begin{tabular}{|l||c|c|r|}
    \hline
    $r_{+}$         & $a$       & $m$  & $l$        \\ \hline\hline
    1               & 0         & 1    & 1          \\ \hline
    0.5 (extremal)  & 0.742     & 1    & 1          \\ \hline
    0.5             & 0.2       & 0.367& 1          \\ \hline
    1               & 0.4       & 0.728& 2 \\ \hline
    1               & 1         & 2    & 1 \\ \hline
  \end{tabular}
\end{table}

\begin{figure}[h!]
 \begin{center}$
 \begin{array}{cccc}
\includegraphics[width=50 mm]{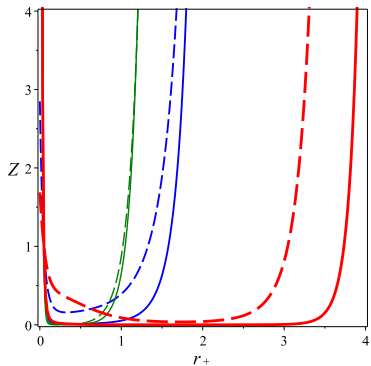}
 \end{array}$
 \end{center}
\caption{Partition function in terms of $r_{+}$ with $l=1$, and $a=0.5$ (blue); $l=1$, and $a=0.742$ (green); $l=2$, and $a=0.4$ (red). Solid and dashed lines represent $\alpha=1$ and $\alpha=0$ respectively.}
 \label{fig:10}
\end{figure}

Furthermore, the temperature $T$ is given by,
\begin{equation}\label{A66}
T=\frac{1}{2\pi r_{+}}\left[\frac{(a^{2}+3r_{+}^{2})(r_{+}^{2}+l^{2})}{2l^{2}(a^{2}+r_{+}^{2})}-1\right].
\end{equation}
The ordinary entropy of the black hole is given by   $s=\frac{A}{4}$, and hence
\begin{equation}\label{A77}
s=\pi\frac{r_{+}^{2}+a^{2}}{\Xi}.
\end{equation}
The thermodynamic volume is given by,
\begin{equation}\label{A1010}
V=\frac{4\pi r_{+}}{3}\frac{r_{+}^{2}+a^{2}}{\Xi}\left(1+\frac{(r_{+}^{2}+l^{2})a^{2}}{2l^{2}r_{+}^{2}\Xi}\right).
\end{equation}
We will now   study logarithmic corrections to the  thermodynamics of this Kerr-AdS black hole.
In the Fig. \ref{fig:10}, we can see behavior of the partition function for various values of $a$ and $l$. We can see
that logarithmic correction reduces value of the partition function. We can reproduce behavior of $Z$, using the following function at the region of $r\leq2$,
\begin{equation}\label{E1}
Z\approx c_{1}+\frac{(c_{2}-c_{3})(r_{+}-b)^{4}}{r_{+}}+\frac{c_{4}(r_{+}-b)}{(r_{+}-d)^{2}},
\end{equation}
where $c_{1}$, $c_{2}$, $c_{3}$ ($c_{2}>c_{3}$), $c_{4}$, $b$ and $d$ are
constants which can fit above function corresponding to the curves of the Fig. \ref{fig:2}. Here $c_{1}=0$
corresponds to the corrected partition function ($\alpha\neq0$),  while $c_{1}\neq0$ and $c_{3}=0$ corresponds to the ordinary entropy ($\alpha=0$).

\begin{figure}[h!]
 \begin{center}$
 \begin{array}{cccc}
\includegraphics[width=50 mm]{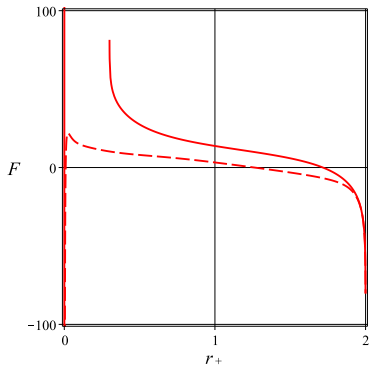}
 \end{array}$
 \end{center}
\caption{Helmholtz free energy in terms of $r_{+}$ with $l=1$, and $a=0.2$. Solid and dashed lines represent logarithmic corrected and ordinary free energy respectively.}
 \label{fig:11}
\end{figure}

In the Fig. \ref{fig:11} we can see typical behavior of free energy $F$ as a function of $r_{+}$. It is illustrated that
the effect of thermal fluctuations increases as  $r_{+}$ becomes smaller. It is clear that logarithmic correction increases Helmholtz free energy.

\begin{figure}[h!]
 \begin{center}$
 \begin{array}{cccc}
\includegraphics[width=50 mm]{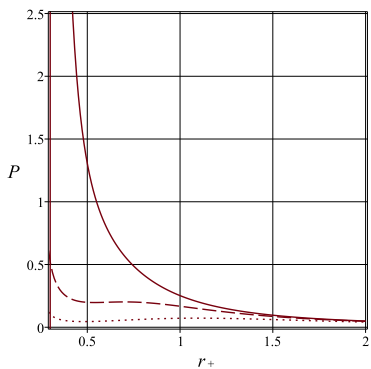}
 \end{array}$
 \end{center}
\caption{Pressure in terms of $r_{+}$ with $l=1$, and $a=0.5$. Solid ($P>P_{C}$), Dashed ($P=P_{C}$), Dotted ($P<P_{C}$).}
 \label{fig:12}
\end{figure}

Now, by using the equation (\ref{P7}) one can discuss about pressure,
and critical point, which are obtained by solving $\partial_{V}P=\partial_{VV}P=0$.
In the Fig. \ref{fig:12}, we can see behavior of pressure compare it with the  critical pressure $P_{C}$.

\section{Conclusions and Discussion}
In this paper, we analyzed the effects of thermal fluctuations on the thermodynamics of a small singly spinning AdS-Kerr black hole in higher dimensional space-time.
It was observed that the entropy of this black hole gets corrected by a logarithmic correction term  due to the thermal fluctuations, and this has important
consequences for the properties of such a black hole. We also calculated the corrections to the various thermodynamical quantities due to the thermal fluctuations.
Then, we analyzed the stability of this black hole in higher dimensions, and the effect of thermal fluctuations on the stability of these black hole. We also
studied the critical  phenomena for such black hole, and the effect of thermal fluctuations on critical phenomena. It was observed that the effect of these thermal
fluctuations can be neglected for a large black hole, but it become important for a sufficiently small black hole. This was expected, as the correction to the
entropy-area relation occurs due to quantum fluctuations in the geometry, and these fluctuations become important only at small scale.\\
This was done studding the phase transition in canonical ensemble. The curve of specific heat for this system were also analyzed, and it was observed that they
have two divergent points (see for example the third plot of the Fig. \ref{fig:4} corresponding to $d=6$).  Furthermore, it was also observed that in four
dimensions    both the large radius region and the small radius region are thermodynamically stable, with positive specific heat. However, the medium radius region are
is unstable with negative specific heat. For the case of $d\leq5$ it is clear that when the size of black hole is small, logarithmic correction make the black hole
unstable. However in higher dimension situation is the inverse.\\
It may be noted that there are several interesting asymptomatic geometries, and it is possible to study the correction to the thermodynamics of such geometries
by thermal fluctuations. It would also be interesting to analyze the effect of thermal fluctuations on black holes in modified theories of gravity.
It is expected that these thermal fluctuations will also correct the  entropy of modified theories of gravity. Such correction to the entropy of modified
theories of gravity, will also produce correction terms for other thermodynamical quantities. This can also affect the critical phenomena which can be studied for
AdS black holes in modified theories of gravity. It may be noted that the critical phenomena has been studied for AdS black holes in $f(R)$ theories of
gravity \cite{adsads1}.\\
The topological black hole solutions of third order Lovelock gravity couple with two classes of Born-Infeld type
nonlinear electrodynamics have been also been studied \cite{adsads2}.
In this analysis, the  geometric and thermodynamics properties of AdS black hole solutions was discussed,
and it was observed that the first law of thermodynamics  holds for such solutions.
The  heat capacity and determinant of Hessian matrix for these black holes was used
evaluate thermal stability in both canonical and grand canonical ensembles. It would be interesting to analyze the effect of thermal fluctuations
on the thermodynamics and stability of such black in Lovelock gravity couple with two classes of Born-Infeld type
nonlinear electrodynamics.\\
It may be noted that recently the   correction to the thermodynamics of black holes has also been obtained from gravity's rainbow \cite{gr12,gr12ab, gr14, gr15, gr17}.
It has been argued that the correction from gravity's rainbow can help resolve the black hole information paradox \cite{info1, infor4, info2}.
It would be interesting to analyze the effects of gravity's rainbow on the thermodynamics of singly spinning AdS-Kerr black hole.
In fact, it is possible to use the results of this paper to obtain certain constraints on the rainbow functions. This is because as the
logarithmic correction is so universally produced by almost all approaches to quantum gravity, we expect that such corrections should also
be produced by gravity's rainbow. This can be used to constrain the parameters used in the rainbow function.
It would also be interesting to perform such an analysis for other kind of black holes, and analyze the relation between correction obtained
from thermal fluctuations and corrections obtained from gravity's rainbow.
It may be noted that   the thermodynamics of black holes has also been studied in massive gravity \cite{mass1, mass2, mass4, mass6}.
It would also be interesting to analyze the effects of thermal fluctuations on the thermodynamics of black holes in massive gravity.
Finally, it may be interesting to study logarithmic correction effect on the $3D$ hairy black hole, as such systems have also been studied \cite{001,002,003,004}. It would  be interesting to analyze the effect that  logarithmic corrections can have on such black holes.

\end{document}